\documentclass{sig-alternate-10pt}
\usepackage{graphicx}
\usepackage[justification=centering]{caption}
\usepackage[format=hang]{subcaption}
\usepackage{url}
\usepackage{amsfonts}
\usepackage{times}
\usepackage{threeparttable}
\usepackage{amsmath, amssymb}
\usepackage{pslatex}
\usepackage{enumerate}
\usepackage{url}
\usepackage{arcs}
\usepackage{yhmath}
\usepackage{epstopdf}

\usepackage[toc,page]{appendix}
\usepackage{color}

\usepackage{booktabs}

\usepackage[ruled,vlined]{algorithm2e}
\usepackage{multirow,array}

\usepackage[nocompress]{cite}
\SetAlFnt{\fontsize{8.75}{8.75}}
\usepackage{hyphenat}

\usepackage{enumitem}

\usepackage{cleveref}
\usepackage{authblk}

\mathchardef\Gamma="0100 \mathchardef\Delta="0101
\mathchardef\Theta="0102 \mathchardef\Lambda="0103
\mathchardef\Xi="0104 \mathchardef\Pi="0105
\mathchardef\Sigma="0106 \mathchardef\Upsilon="0107
\mathchardef\Phi="0108 \mathchardef\Psi="0109
\mathchardef\Omega="010A

\newcommand{\outline}[1]{}

\usepackage{cite}
\usepackage{xspace}
\usepackage{url}
\usepackage{graphicx}
\usepackage{latexsym}
\usepackage{amssymb}
\usepackage{amsfonts}
\usepackage{psfrag}
\usepackage{wrapfig}
\usepackage{comment}
\usepackage{alltt}
\usepackage{color}

\newcommand{\etal}{\frenchspacing{}\emph{et al{.}}\xspace}

\newcommand{\Comment}[1]{}

\renewcommand{\color}[1]{}

\newcommand{\qns}{quantum networks\xspace}
\newcommand{\qn}{quantum network\xspace}
\newcommand{\QE}{Quantum entanglement\xspace}
\newcommand{\qe}{quantum entanglement\xspace}
\newcommand{\qes}{quantum entanglements\xspace}

\begin{document}

\date{}

\title{\Large \bf Modeling and Designing Routing Protocols in Quantum Networks}

\author[1]{Shouqian Shi}

\author[1]{Chen Qian}

\affil[1]{University of California, Santa Cruz}

\pagenumbering{arabic}
\maketitle

\subsection*{Abstract}
Quantum networks enable a number of important applications such as quantum key distribution. The basic function of a quantum network is to enable long-distance \qe between two remote communication parties. This work focuses on the entanglement routing problem, whose objective is to build long-distance entanglements for the concurrent source-destination pairs through multiple hops. Different from existing works that analyzes the traditional routing techniques on special network topologies, we present a comprehensive entanglement routing model that reflects the differences between quantum networks and classical networks and  new entanglement routing algorithms that utilize the unique properties of quantum networks. Evaluation results show that the proposed algorithm Q-CAST increases the number of successful long-distance entanglements by a big margin compared to other methods. The model and simulator developed by this work may encourage more network researchers to study the entanglement routing problem.

\section{Introduction}

A quantum network (also called as a quantum Internet) is an interconnection of quantum processors and repeaters that can generate, exchange, and process quantum information \cite{QuantumInternet2008,QuantumInternet2018,QuantumInternet2018_2,QuantumInternet2019}. It facilitates the transmission of information in the form of quantum bits, also called \emph{qubits}, between physically separated quantum memory.
Long-distance quantum information exchange has been proposed, studied, and validated since 1980s \cite{BB84,E91,Pan98,TokyoQKD,DARPAQuantum,Vienna,Satellite} and many experimental studies have demonstrated that communication of quantum information can become successful in reality, such as  the DARPA quantum network \cite{DARPAQuantum}, SECOQC Vienna QKD network \cite{Vienna}, the Tokyo QKD network \cite{TokyoQKD}, and the satellite quantum network in China \cite{Satellite}.

Quantum networks are not meant to replace the classical Internet communication. In fact, they supplement the classical Internet and enable a number of important applications such as quantum key distribution (QKD) \cite{BB84,E91,pir2019advances}, clock synchronization \cite{QuantumClock}, secure remote computation \cite{remotequantum}, and distributed consensus \cite{distributedQuantum}, most of which cannot be easily achieved by the classical Internet.

The basic function of a quantum network is to enable long-distance \qe between two remote communication parties. Hence, most applications of \qns are developed based on two important features of \qe. 1) Quantum entanglements are inherently private by the laws of quantum mechanics such as the ``no-cloning theorem'' \cite{nocloning} and hence prevent a third party from eavesdropping the communication \cite{E91}.

\qe is a perfect solution of the most fundamental problem of network security: key distribution (also known as key agreement) \cite{DiffieHellman}. Compared to public key cryptography \cite{RSA}, quantum key distribution (QKD) has provable security based on information theory and forward secrecy \cite{QuantumInternet2018}, instead of relying on the computational complexity of certain functions (such as factorization). 2) \QE provides strong correlation and instantaneous coordination of the communication parties.
 Hence, \qe can achieve tasks that are difficult to coordinate in classical networks due to unexpected network latencies, such as clock synchronization \cite{QuantumClock} and distributed consensus \cite{distributedQuantum}.

Recent progress reveals that \qns could become practical in 5 years \cite{QuantumInternet2018}, and they do not rely on the well-functioning quantum computers carrying a sufficient amount of qubits. In fact, many applications of \qns can be implemented with one or two qubits. Considering the QKD example, we are able to distribute a secret bit with only one entanglement pair. By repeating the 1-pair QKD process we can generate secret keys with a sufficient length. Hence, research on \qns is a timing topic.

To generate a long-distance \qe between two parties Alice and Bob, one of them should create an entangled pair of photons and send one photon to the other party through a channel using certain physical media such as optical fiber. However, optical fiber is inherently lossy and the success rate $p$ of establishing an entanglement pair decays exponentially with the physical distance between the two parties \cite{pirandola2009direct,repeaterless2014}. Hence, to increase success rate of long-distance \qe, a number of \emph{quantum repeaters} need to be deployed between two long-distance communication parties \cite{QuantumInternet2018,repeaterless2014}. Eventually a network of quantum repeaters will be deployed to support any-to-any communication of world-wide quantum processors, similar to the evolution of classical Internet. One critical issue is that quantum repeaters work in \emph{a completely different way} from classical network routers: they use quantum swapping instead of the packet switching. Hence, new algorithms are required to be designed and how to reliably generate \qe using the network of repeaters remains an unsolved yet important problem.

This work focuses on a key problem called \emph{entanglement routing}, whose objective is to build long-distance entanglements through multiple hops for any pair of source and destination in the network.
This problem  can be considered on the network layer of a quantum network \cite{LinkLayerQuantum}.
Existing work that investigate the routing problem of quantum networks is limited to analyzing the traditional routing techniques (Dijkstra shortest paths, multipath routing, and greedy routing) on special network topologies (ring, sphere, or grid), such as the very recent ones \cite{RoutingEntanglement,GreedyRoutingQuantum}.
In this study, we present a comprehensive entanglement routing model that reflects the difference between quantum networks and classical networks and  new entanglement routing designs that utilize the unique properties of quantum networks. The proposed algorithms include realistic protocol-design consideration such as  arbitrary network topologies, multiple concurrent sources and destinations to compete resource, link state exchanges, and limited qubit capacity of each node, most of which have not been considered by prior studies.

Evaluation results show that the proposed algorithm Q-CAST increases the number of successful long-distance entanglements by a big margin compared to other methods.
More importantly, this study may encourage more network researchers to  study the entanglement routing problem. We present and clarify the models and problems of entanglement routing, with the comparison of similar terms and concepts used in classical network research.
A simulator with algorithm implementation, topology generation, statistics, and network visualization functions is built and will be open to public  \cite{QuantumCode}.

The rest of this paper is organized as follows. \S~\ref{sec:related} presents the related work of quantum network routing and \S~\ref{sec:model} introduces the network model. We present the algorithm designs in \S~\ref{sec:design}.

The evaluation results are shown in \S~\ref{sec:eva}. We discuss some related issues in \S~\ref{sec:discuss} and conclude this work in \S~\ref{sec:conclusion}.

\vspace{-3ex}
\section{Related Work}
\vspace{-1.5ex}
\label{sec:related}

Quantum information exchange has been proposed, studied, and validated for more than 20 years \cite{BB84,E91,Pan98,TokyoQKD,DARPAQuantum,Vienna,Satellite}.
The concept of quantum networks is first introduced by the DARPA quantum network project aiming to implement secure communication in the early 2000s \cite{DARPAQuantum}. This project uses multiple physical layer implements including fiber optics to transmit entangled photons. 
Recent implementations include the SECOQC Vienna QKD network \cite{Vienna}, the Tokyo QKD network \cite{TokyoQKD} and China's satellite quantum network \cite{Satellite}.

In addition to these experimental work, researchers have started the algorithm designs of quantum networks that can provide reliable and fast services to quantum network applications. One fundamental problem is to routing quantum entanglements with reliability protection using the quantum repeater network with lossy links \cite{repeater2013}. Pirandola \emph{et al.} discuss the limits of repeaterless quantum communication \cite{repeaterless2014} and discuss the multipath routing in a diamond topology \cite{pirandola2019end}. Schoute \emph{et al.} \cite{Shortcuts} proposed a framework to study quantum network routing. However, their discussion is only limited to ring or sphere topology. Das  \emph{et al.} \cite{RobustQuantum} compares different special topologies for entanglement routing. Caleffi \cite{optimalQuantum} studies the optimal routing problem in a chain of repeaters. Pant \emph{et al.} \cite{RoutingEntanglement} proposes solutions for entanglement routing in grid networks. \cite{GreedyRoutingQuantum} proposes virtual-path based greedy routing in ring and grid networks. For all studies mentioned above, they assume specialized network topologies, which may be over-simplified. The topologies of practical quantum networks may be arbitrary graphs because 1) the end hosts in quantum networks (i.e., trusted nodes) must exist on specified locations according to application requirements, instead of following certain topologies; 2) deploying unnecessary trusted nodes and quantum repeaters just to create certain topologies is a waste of resource.
The simple greedy routing algorithms used in \cite{RoutingEntanglement,GreedyRoutingQuantum} may fail at the local minimums in arbitrary graphs as being extensively studied in prior research for wireless routing \cite{GPSR,MDT}.
In addition, existing methods are limited to study a single pair of source and destination and do not consider concurrent source-destination pairs that exist in most cases of the practice.

Recently
Dahlberg \etal \cite{LinkLayerQuantum} give a reference model of quantum network stack, which contains the physical layer, link layer, network layer, and transport layer. Based on that, they provide a reliable physical and link layer protocol for quantum networks on the NV hardware platform. 
The routing algorithms proposed in this paper fit in the network layer to provide the concurrent entanglement routing solutions.

\section{Network Model and Problem Definition}
\label{sec:model}

The network model used in this study follows the facts from existing physical experiments \cite{Pan98,qubit2013,teleportation2009,entanglement2007} and the corresponding studies \cite{Shortcuts,RobustQuantum,RoutingEntanglement} to reflect a practical \qn.
Compared to prior models used in existing studies of quantum network performance \cite{Shortcuts,RobustQuantum,RoutingEntanglement,GreedyRoutingQuantum} -- mostly by physicists and theoreticians -- our model includes many practical considerations and experience from network protocol designs, such as the  dynamics of quantum links, definition and comparison of different routing metrics, concurrent source-destination pairs,

limited qubit capacity of each node, clear differentiation of the network topology and link state information, and  limited link-state propagation in a time slot.

\subsection{Network Model}

There are three main components in a quantum network \cite{stefano2016unite,QuantumInternet2018}, explained as follows.

\textbf{1. Quantum processors.} Quantum processors, also called `trusted nodes' \cite{Shortcuts,RoutingEntanglement}, are similar to the end hosts in classical networks, which run the network applications that require end hosts to communicate with each other. Different from classical end hosts, each quantum processor is equipped with a certain size of \emph{quantum memory}, quantified in the number of qubits. All quantum processors are connected via the classical Internet. Hence, they are able to freely exchange classical information. \emph{Entanglement routing} requires the \qn to establish an entanglement between two target quantum processors through a number of quantum links, which may be simply called `routing' in existing work \cite{RobustQuantum,RoutingEntanglement}.

\textbf{2. Quantum repeaters.} It is difficult to establish a shared entanglement between two distant quantum processors, because the fidelity of entanglements decay exponentially with the distance.  Hence, quantum repeaters are used as relays to connect quantum processors. They support entanglement sharing over arbitrarily long distances via \emph{quantum swapping} (explained later). A quantum repeater may also connect to other repeaters and quantum processors via the classical Internet to exchange the control messages.
Every quantum processor also includes the complete function of a repeater.

    Compared to a quantum processor, a repeater \emph{cannot} be the source or destination of  any qubit transmission -- it can only perform quantum swapping but not measurement.
   Quantum processors and repeaters are both called \emph{quantum devices} or simply \emph{nodes}.

\textbf{3. Quantum channels and links.}
A quantum channel connecting two quantum devices supports the transmission of qubits. The physical material of quantum channels may be standard optical fibers. A quantum channel is inherently lossy: the success rate of each attempt to create an entanglement of a quantum channel $c$ is $p_c$, which decreases exponentially with the physical length of the channel: $p_c \sim e^{-\alpha L}$\footnote{The success rate of a link is  determined by the physical layer and link layer, taking into account the channel transmissivity, fidelity of transmitted entanglements, number of permitted entanglement trials in one phase, and the link layer algorithm \cite{LinkLayerQuantum,RoutingEntanglement}. In the link layer, a channel is allowed multiple attempts to build a link, and the link is established on the first successful attempt. The $p_0$ here is the overall success rate. } where $L$ is the physical length of the channel
 and $\alpha$ is a constant depending on the physical media \cite{QuantumInternet2018,RoutingEntanglement,rate-loss2014,repeaterless2014}.
If an entanglement attempt is successful, we call there is a \emph{quantum link} on this channel and the two quantum processors share an entanglement pair.

\begin{figure}
 \centering
 \includegraphics[width=0.9\linewidth]{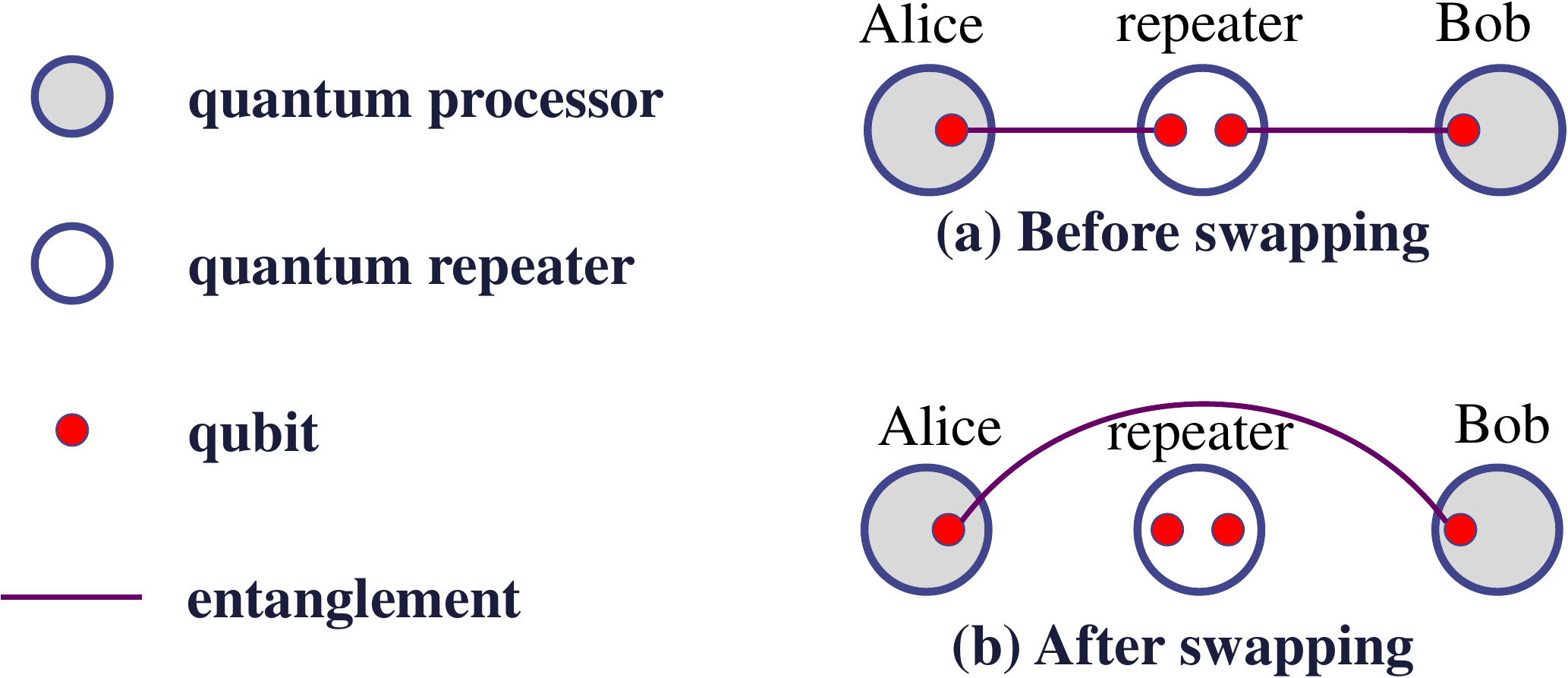}

 \caption{Quantum swapping to establish a long-distance entanglement. }
 \label{fig:swapping}

\end{figure}

\textbf{Quantum swapping.} Classical forwarding devices, such as network routers, perform packet switching in a store-and-forward manner.

A quantum repeater works in \emph{a completely different way}. As shown in Figure \ref{fig:swapping}(a), to create an entanglement pair for two distant quantum processors Alice and Bob, a quantum repeater first generates an entanglement pair with Alice and simultaneously generates another entanglement pair with Bob.

After performing the Bell state measurement (BSM) \cite{Pan98,teleportation2009,qubit2013}, the repeater can teleport the qubit entangled with Alice to Bob, resulting an entangled qubit pair, also called an \emph{entangled qubit} (\emph{ebit}), shared by Alice and Bob, as shown in Figure \ref{fig:swapping}(b). Hence, Alice and Bob can share information via the ebit. This process is called \emph{quantum swapping} (or quantum teleportation) \cite{teleportation2009,QuantumInternet2018,RoutingEntanglement}. Similar operations can be performed by multiple repeaters on a path connecting Alice and Bob. Unlike the hop-by-hop forwarding in classical networks, quantum swapping requires all quantum links on the path to be successful \emph{at the same time}.

The entanglement routing problem is to establish an ebit between two given quantum processors, via quantum swapping, in a quantum network. It is the fundamental service of the \qn for  applications of two remote users, such as QKD.

\textbf{Network topology.}
As a formal specification, we consider a network of nodes described by a weighted multigraph $G=\langle V, E, C \rangle$. $V$ is the set of $n$ nodes. Each node $u$ is either a quantum processor or repeater, equipped with a limited number $Q_u$ of qubits to build quantum links.
A quantum processor can also perform quantum swapping for neighboring nodes and only a quantum processor can be the source or destination of qubit transmission.
We assume all nodes are connected via classical networks, i.e., the Internet, and every node has a certain level of classical computing and storage capacity, such as a desktop server.

\textbf{Edges, channels, links, and paths.}
$E$ is the set of edges in the graph. An edge existing between two nodes means that the two nodes shares one or more quantum channels. The number of channels $W$ is called the \textit{width} of the edge.
$C$ is the set of all channels, each of which is identified by its two end nodes.

Quantum processors can assign different memory qubits to different quantum channels. Channels that are assigned qubits at both ends are \textit{bound channels}, other channels are \textit{unbound channels}. There could be multiple parallel channels between two nodes. At a given time two neighbor nodes may share multiple quantum links.

In order to create direct \qe, two neighbor nodes must make \qe attempts in the same time on the shared bound channels. On the other hand, two nodes share a \textit{quantum link} means they have a direct \qe at this moment.

A path between a pair of nodes is identified by the sequence of the nodes along the path $v_0, v_1, \cdots, v_h$, and the \textit{width $W$} of the path. A path has width $W$ means each edge of the path has at least $W$ parallel channels. For succinctness, the path $\langle (v_0, v_1, \cdots, v_h), W\rangle$ is also called a $(W,h)$-path, or a $W$-path. As a different concept, if an end-to-end path of quantum links (not channels) that can build $w$ entanglements, i.e., each edge has at least $w$ successful quantum links, we call the path is a \textit{$w$-entangled path}, or the path has entanglement width $w$.

\begin{figure}[t]
 \begin{center}
   \includegraphics[width=1\linewidth]{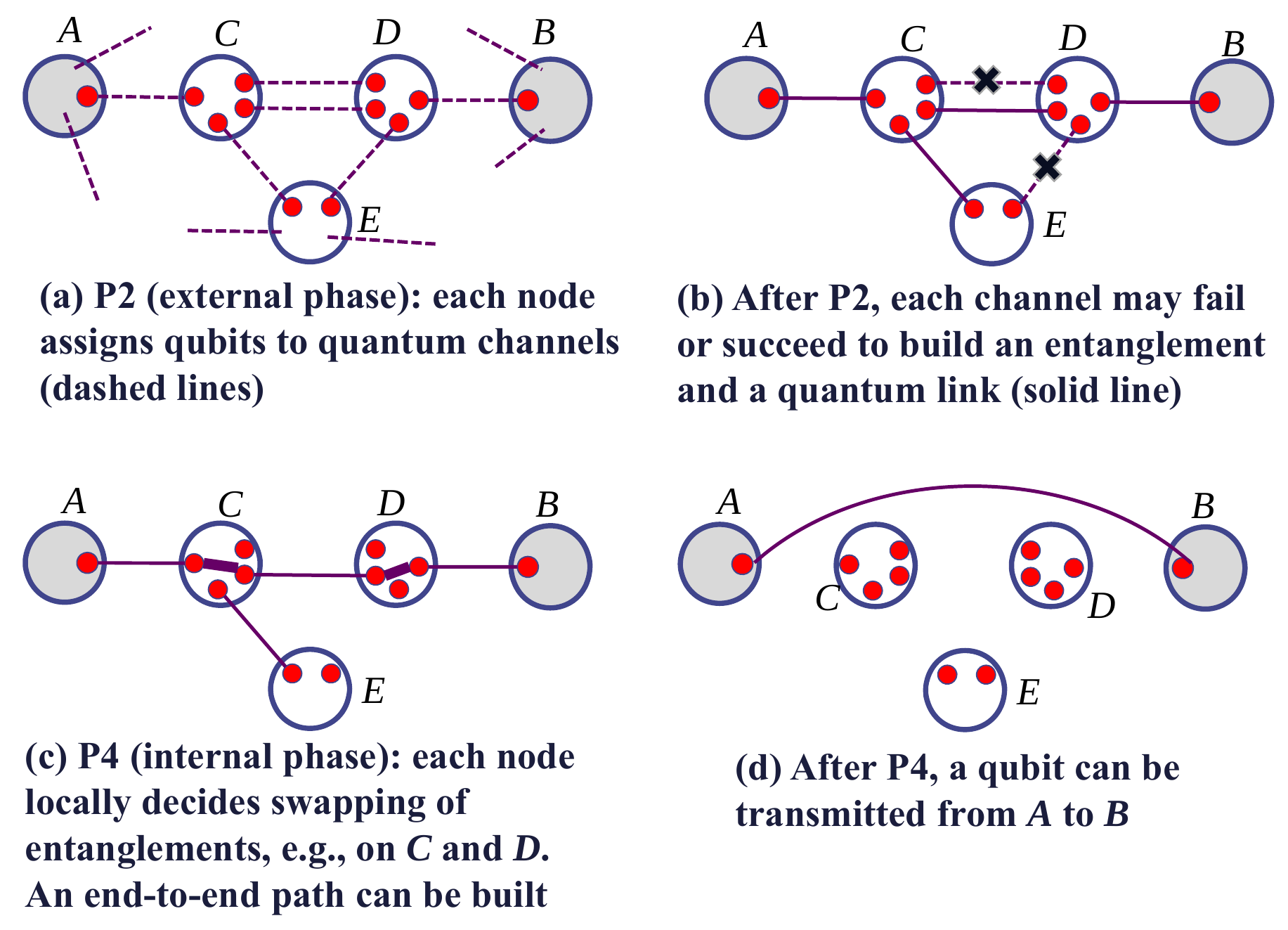}
 \end{center}

 \caption{\small Phases 2 and 4 of a time slot. Entanglement routing aims to build an end-to-end path with successful links after the time slot. }
 \label{fig:timeslot}

\end{figure}

\textbf{Time slots.}

For long-distance quantum swapping, all nodes on the path need to obtain the \qe with its predecessor and successor at the same time. Hence, some level of time synchronization among all nodes is necessary, which can be achieved by the current synchronization protocols via Internet connections.  Time is loosely synchronized in \emph{time slots} \cite{RoutingEntanglement}. Each time slot is a device-technology-dependent constant  and set to an appropriate duration by the link layer

such that the established entanglements do not discohere within one time slot \cite{RoutingEntanglement}.

Before each time lost, each node should know the global network topology $G=\langle V, E, C \rangle$, which is relatively stable.

Each time slot includes four phases as an extended model from \cite{RoutingEntanglement}. In Phase One (\textbf{P1}), all nodes receive the information of current source-destination (S-D) pairs that need to establish long-distance entanglements by the Internet.

Phase Two (\textbf{P2}) is called the external phase  \cite{RoutingEntanglement}. In P2, each node assigns its quantum memory qubits to quantum channels and attempts to generate \qes with neighboring nodes on the bound channels.
As an example in Figure \ref{fig:timeslot}(a), $A$ and $B$ are distant quantum processors that wants to exchange some qubit information. $C$, $D$, and $E$ are repeaters. Each node (processor or repeater) has a number of qubits (red dots) for quantum networking purposes.

The dashed lines are quantum channels.  Two neighbors may share multiple channels

simultaneously. Because qubits are limited resource, some channels are not assigned qubits and are not used in this time slot.
During P2, each channel can make a finite number $n_c$ attempts \cite{DeterministicDelivery}, $n_c\geq1$, until a link is built or P2 finishes.
After P2, some quantum links may be created as shown in Fig.~\ref{fig:timeslot}(b). We call the information of these links as \emph{link states}. Compared to the same term in link-state routing of classical networks \cite{OSPF}, the quantum link states are \emph{highly dynamic and nondeterministic}. In Phase Three (\textbf{P3}), each node
knows its own link states via classical communications with its neighbors \cite{RoutingEntanglement} and shares its link states via the classical network.
Since successful entanglements will quickly decay, each node can only  receive the link states of a subset of other nodes.
P3 only includes classical information exchange.
In Phase Four (\textbf{P4}), also called the internal phase  \cite{RoutingEntanglement}, nodes perform quantum swapping to establish long-distance quantum entanglement using the successful quantum links. Each node \emph{locally} determines the swapping of successful entanglements, which can be considered as placing an \emph{internal link} between two qubits as shown in Fig.~\ref{fig:timeslot}(c).
Each swapping succeeds at a device-dependent probability $q$. $A$ and $B$ can successfully share an entanglement qubit pair (\textit{an ebit}) if there is an end-to-end path with both external and internal links as  in Fig.~\ref{fig:timeslot}(c).

\textbf{Local link-state knowledge.}
P3 and P4 should be short such that the successful entanglements do not decay. Hence it is impractical for a node to know the global link states within such short time as the classical network has latencies \cite{RoutingEntanglement}. Hence, a practical model is to allow each node to know the link states of its $k$-hop neighbors, $k\geq1$ \cite{GreedyRoutingQuantum}. The swapping decisions in P4 thus include the $k$-hop link-state information as the input. It is obvious that the routing path selection could be sub-optimal without global link-state knowledge.

\textbf{Exclusive qubit/channel reservation.}
In P2 of each time slot, to establish a single link on a channel, each end of the channel is assigned a qubit. This qubit-channel assignment is exclusive: the qubits cannot be shared by other channels, and no more qubits can be assigned to the channel. And in P4, to generate an ebit shared by a pair of distant nodes, quantum swapping is performed on pairs of links. This quantum swapping is also exclusive, and a single link cannot be used for more than one swapping. Hence, the qubits and channels are precious \textit{routing resource} and should be carefully managed.

\subsection{The entanglement routing problem}

This work studies the entanglement routing problem: we are given a quantum network with an arbitrary network graph $G=\langle V, E, C \rangle$ and a number of source-destination (S-D) pairs $(\langle s_1, d_1\rangle, \langle s_2, d_2\rangle, \cdots, \langle s_m, d_m\rangle)$, where $s_i,d_i \in V$. The number of memory qubits of a node $u \in V$ is $Q_u$, and each edge $e \in E$ consists of one or more channels from $C$. For each bound channel $c$, a link is successfully built in probability $p_c$ in P2. In P3, each node gets the link-state information of its $k$-hop neighbors. Each node decides the swapping of its internal qubits in P4 \emph{locally}, and each swapping succeeds in probability $q$.

The objective of entanglement routing is to \emph{maximize the number of ebits} delivered for all S-D pairs, in each time slot. Each ebit must be delivered by a long-distance quantum entanglement, built by a path of successful quantum links from the source to the destination. Each S-D pair may share multiple ebits.
 A successful quantum link can only be used for one long-distance entanglement.
The number of ebits for one S-D pair in one time slot is also called the \textit{throughput} between the S-D pair. The objective can then be set to maximize the overall throughput in the network.

Note that this objective does not consider fairness among different S-D pairs, but we show the proposed algorithms achieve a certain level of  fairness as in \S~\ref{sec:eva}. In addition, in \S~\ref{sec:discuss} we propose a simple extension to our designs to achieve better fairness among S-D pairs.

\subsection{Compared to classical network routing}

We summarize the differences between quantum entanglement routing and classical network routing. We show that existing routing techniques are not sufficient to solve the entanglement routing problem.

\textbf{Term clarification.}  Edges, channels, and links have \emph{different} definitions as presented above in this model, although they are used interchangeably in classical networks. Besides, the network topology and global link states may be considered as similar information in classical routing such as OSPF \cite{OSPF}. However, in a quantum network, while the network topology (nodes and channels) is stable and known to all nodes, the link states (whether the entanglements succeeded) are dynamic and only shared locally in P3 and P4 of each time slot.

\textbf{Versus routing in wired packet-switching networks.} Link-state and distance-vector are two main types of
routing protocols for packet-switching networks. Main differences: \textbf{1)} Packet switching relies on either link-state broadcast or multi-round distance vector exchanges to compute the shortest paths. However, in a quantum network, link states are local information and a node may not know whether a link that is $k$-hop away is successfully built or not. There is no time for global link-state broadcast or distance vector convergence, because entanglements on the links will quickly decay. \textbf{2)} Quantum links are highly unreliable while wired links are relatively reliable.  \textbf{3)} A quantum link cannot be shared by multiple S-D pairs, which is allowed in classical packet switching. If a link is claimed by multiple S-D pairs, it can only satisfy one of them. Hence, the ``shortest paths'' computed by classical routing will not always be available. \textbf{4)} Classical packets can be buffered on any node for future transmission. In quantum networks, links on a paths must be successful in the same time slot.

\textbf{Versus routing in multi-hop wireless networks.} A multi-hop wireless network could be either a mobile ad hoc network \cite{adhocrouting} or a wireless sensor network \cite{sensorsurvey}.  Main differences: \textbf{1)} For an ad hoc wireless node, neither the network topology nor global link state is known. For a quantum node, although link state is local information, the network topology is known in advance via the Internet. \textbf{2)} An ad hoc wireless node can keep sending a packet until the transmission is successful or  a preferred receiver moves close to it. Each quantum link can only be used once and all links on an end-to-end path must be available simultaneously. Existing wireless ad hoc routing methods such as DSR \cite{DSR}, AODV \cite{AODV}, and geographic routing \cite{GPSR} are all packet-switching protocols and do not fit quantum networks. Also, they do not take the global network topology information.

\textbf{Versus circuit-switching and flow scheduling in SDN.} Circuit switching, virtual circuit, and flow scheduling in software defined networks (SDNs) all need to pre-determine the end-to-end paths and reserve certain resource on the paths, such as \cite{VCRouting,ATMRouting,RandomizedCircuit,Hedera}, which share similarity with entanglement routing. The main difference is that in a quantum network, links on reserved paths may arbitrarily fail, and hence more robust path allocation and recovery algorithms are required.

\section{Entanglement Routing Algorithms}
\label{sec:design}

The proposed entanglement routing algorithms utilize the unique properties of quantum networks that have not been explored in classical network routing. Compared to recent quantum network studies \cite{Shortcuts,RobustQuantum,RoutingEntanglement,GreedyRoutingQuantum}, the proposed protocols follow a standard protocol-design approach and use more realistic network models: arbitrary network topologies, multiple concurrent S-D pairs to compete links, link state exchanges, and limited qubit capacity of each node.

\subsection{Main ideas}

Our design is based on the following three innovative ideas to utilize the \textbf{unique features} of a quantum network:

\textbf{1. Path computation based on global topology and path recovery based on local link states.}

The quantum network graph $G=\langle V, E, C \rangle$ is relatively stable and hence can be known to every node. However, the link states are highly dynamic and probabilistic in each time slot. The frequent link state changes cannot be propagated throughout the whole network, especially when the entanglements decay quickly.
Hence, nodes select and agree on the same list of paths based on global topology information in P2, and try to recover from link failures based on local link states in P4.

\begin{figure}

	\centering
	\includegraphics[width=0.75\linewidth]{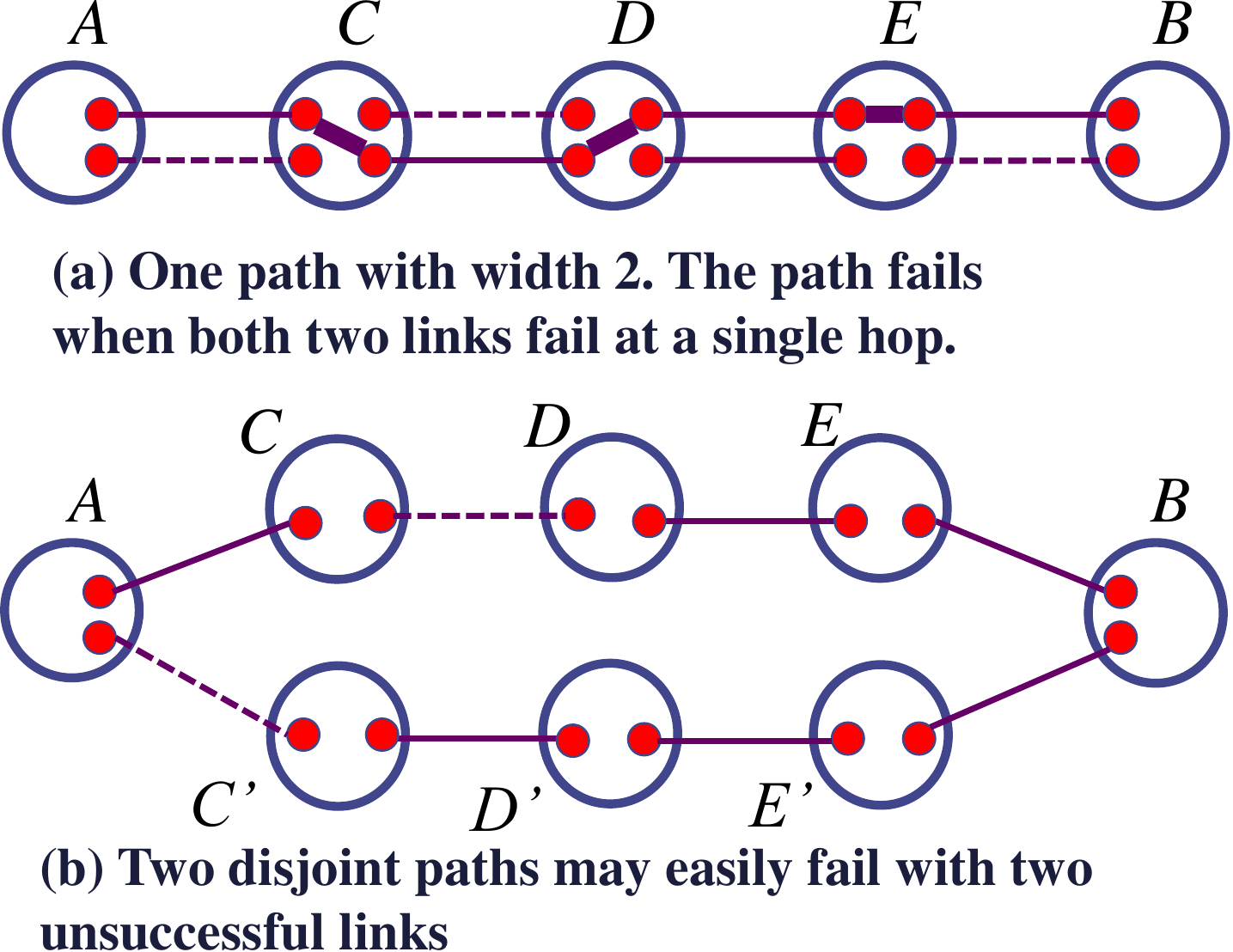}

	\caption{A wide path (subfigure a) is more reliable than disjoint paths (subfigure b) using the same resource}
	\label{fig:parallel}

\end{figure}

\textbf{2. Wide paths are preferred.} Recall that on a $W$-path, each edge has at least $W$ parallel channels. Fig.~\ref{fig:parallel}(a) shows an example of a 2-path from $A$ to $B$. Compared to two disjoint paths shown in Fig.~\ref{fig:parallel}(b), which cost the same amount of resource (qubits and channels), the wide path is more reliable because it only fails when two links fail simultaneously at a single hop. Suppose the success rate of each channel is 0.5. We may easily calculate that the 2-path can build at least one long-distance entanglement with probability 0.32, while this probability for the two disjoint paths is 0.12.
To achieve high throughput on a path with $W>1$, nodes should share a consensus on how to perform swapping (place internal links in Fig.~\ref{fig:parallel}) instead of making choice randomly. Each channel is assigned a global unique ID. During P4, each node places an intern link between the link with the smallest ID to its predecessor and the link with the smallest ID to its successor. And it repeats this process until no intern link can be made for this path.

\begin{figure}[t!]
		\centering\includegraphics[width=.7\linewidth]{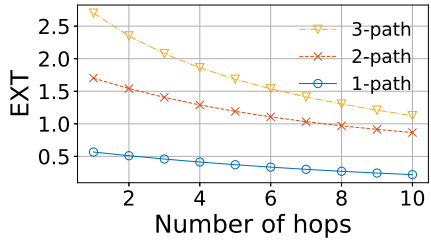}

		\caption{\small EXT, $p=0.9$}
		\label{fig:E-hops-0.9}
\end{figure}

\begin{figure}[t!]
	\centering\includegraphics[width=.7\linewidth]{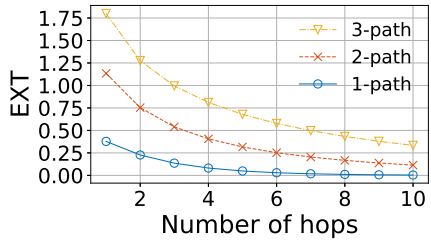}

	\caption{\small EXT, $p=0.6$}
	\label{fig:E-hops-0.6}

\end{figure}

Formally, we may define a \emph{routing metric}, called the expected number of ebits or expected throughput (\textbf{EXT}) $E_t$, to quantify an end-to-end path on the network topology. For a $(h,W)$-path $P$, suppose the channel entanglement success rate on the $i$-th hop is $p_i$, where $i \in \{1, 2, \cdots, h\}$. We denote the probability of \textit{the $k$-th edge on the path having exactly $i$ successful links} as $Q_k^i$, and the probability of \textit{the first $k$ hops of $P$ is an $i$-entanglement path} as $P_k^i$. Then we get the recursive formula set, for $i \in \{1,2,\cdots, W\}$ and $k \in \{1,2,\cdots, h\}$:

	\begin{equation} \label{eq:P}
	\begin{aligned}
	Q_k^i &= C_W^ip_k^i(1-p_k)^{W-i} \\
	P_1^i &= Q_1^i \\
	P_k^i &= P_{k-1}^i \cdot \sum_{l=i}^{W} Q_k^l + Q_k^i \cdot \sum_{l=i+1}^{W} P_{k-1}^l
	\end{aligned}
	\end{equation}

Further, considering the success probability $q$ of each quantum swapping, we get the EXT $E_t = q^h \cdot \sum_{i=1}^{W} i \cdot P_h^i$. We show some numerical results. For simplicity, we set $p_1, p_2, \cdots, p_h = p$, and let $p$ be 0.9 or 0.6. We vary the $W$ from 1 to 3 and the $h$ from 1 to 10, and the results of the EXT of a $W$-path are shown in \Cref{fig:E-hops-0.9,fig:E-hops-0.6}. It obvious that a $W$-path has significant improvement of EXT over a 1-path, for more than a factor of $W$.

\textbf{3. Offline pre-computation versus contention-aware online path selection.} In different time slots, the S-D pairs may be different. We propose two approaches to select paths for S-D pairs in each time slot. The first approach utilizes offline computation, which happens at any time before a time slot, such as during system initialization. Multiple paths for \emph{each} potential S-D pair is pre-computed, and all nodes stores these paths. At each time slot, nodes select the pre-computed paths for current S-D pairs. The contention-aware online algorithm does not pre-compute the paths for all S-D pairs. At each time slot, the algorithm finds contention-free paths for current S-D pairs. A set of paths are `\textit{contention-free}' if the network can simultaneously satisfy the qubit and channel requirement for all the paths in full width. We propose two algorithms using the two approaches, called Q-PASS and Q-CAST.

\subsection{Q-PASS: Pre-computed pAth Selection and Segment-based recovery}
\label{sec:alg1}

\subsubsection{Algorithm overview}

\begin{figure}
\centering
\includegraphics[width=\linewidth]{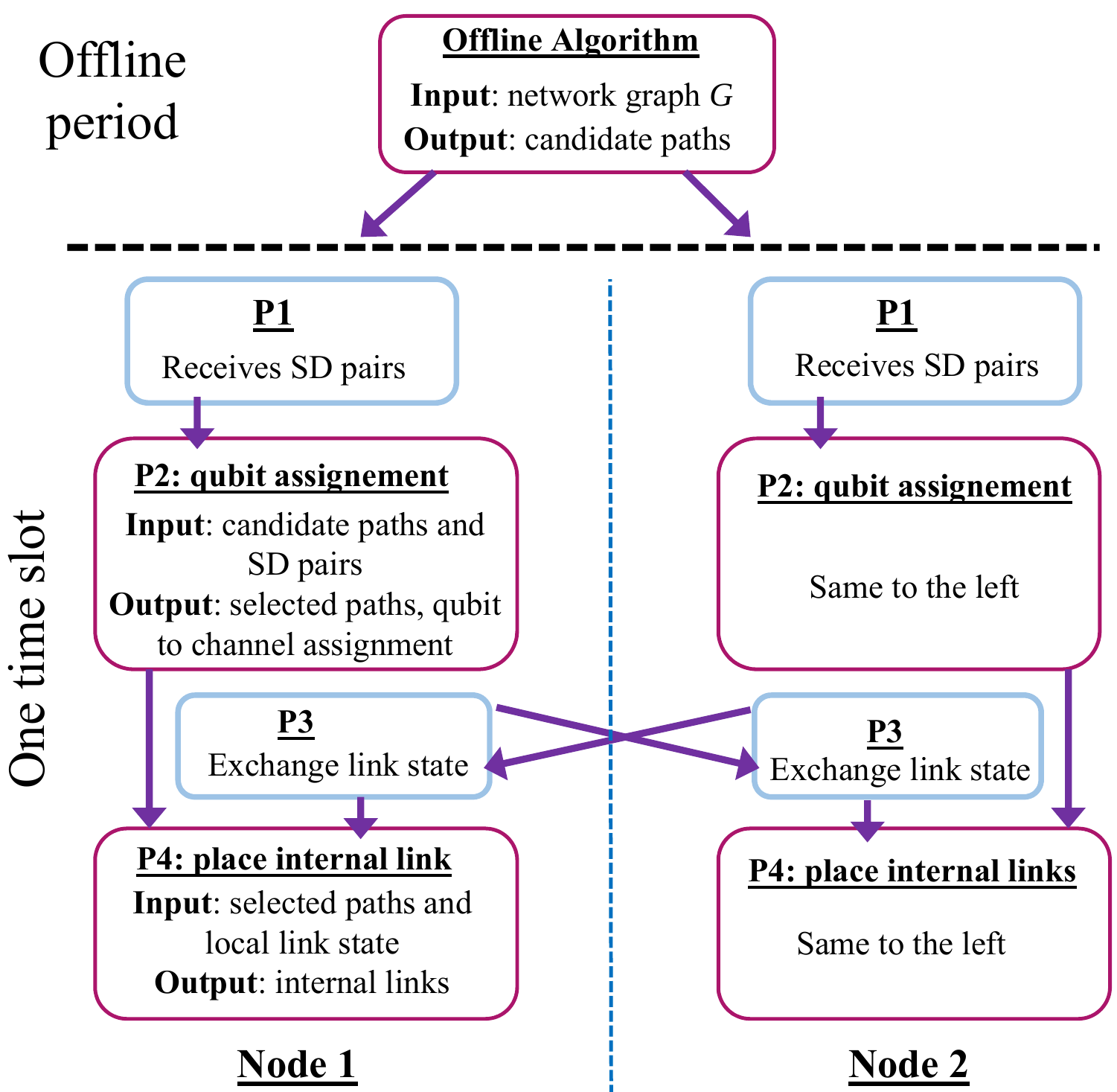}

\caption{The algorithm workflow  of Q-PASS. Nodes 1 and 2 are two arbitrary neighbor nodes and run the same algorithm. }
\label{fig:QPass}

\end{figure}

We present the algorithm Q-PASS, whose workflow is shown in Fig.~\ref{fig:QPass}.

The core idea of Q-PASS is to pre-compute potential `good' paths between \emph{all possible S-D pairs} based on the network topology $G$. Then in each time slot, every node uses an online algorithm to make qubit-to-channel assignments based on the pre-computed paths of \emph{current S-D pairs} and make local swapping decisions based on local link states.
The design includes both offline and online algorithms.

The offline phase (top of Fig.~\ref{fig:QPass}) may happen at the system initialization or after the network topology changes. The results of an offline phase can be used by many succeeding time slots  until a topology change happen.
Hence, we may assume the time for an offline period is sufficiently long.
The offline algorithm is run at a trusted server, with replica servers for robustness. These servers connect to all quantum nodes via classical networks.
The outputs of the offline algorithm are the ``candidate paths'' for all possible S-D pairs. The candidate paths of each S-D pair are those paths connecting the S-D and with small values of the selected metric.

The algorithm of each time slot (bottom half of Fig.~\ref{fig:QPass}) is a four-phase  design and run by each node in a distributed and concurrent manner. It should be fast and only use the $k$-hop link-state information.
P1 and P3 only include standard processes and do not have special algorithmic design. Q-PASS P2 takes the candidate paths from the offline algorithm and the S-D pairs as the input. It computes a number of selected paths for the S-D pairs and its local qubit-to-channel assignment. Note that the inputs are globally consistent on all nodes. Hence, the selected paths are also consistent on different nodes. The assignment will produce a number of successful links in P2. And in P3, nodes exchange the link states with their $k$-hop neighbors. Q-PASS P4 uses the selected paths and link state information as the input to compute the the swapping decisions (i.e., internal links). After P4, possible long-distance entanglement can be built for S-D pairs.

We present the offline, P2, and P4 algorithms of  Q-PASS in details.

\begin{figure*}
\centering
\includegraphics[width=\linewidth]{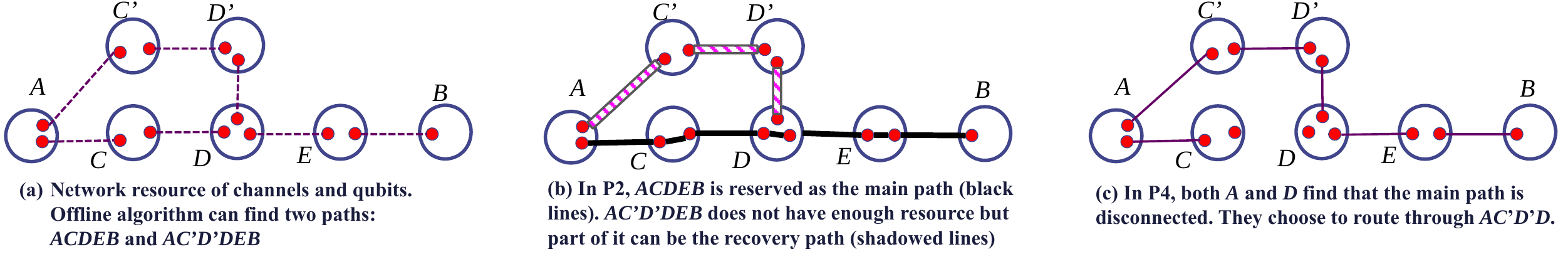}

\caption{Example  of Q-PASS P4. $A$ and $B$ are the S-D pair.}
\label{fig:QPass-algo}

\end{figure*}

\subsubsection{Offline path computation}

The offline algorithm should find multiple paths for each S-D pair to provide multiple potential paths to select  in each time slot. We use Yen's algorithm \cite{kshort} to get multiple paths for each pair. Note that the results of Yen's algorithm is not contention-free: the paths may overlap at nodes or channels, and the network may not have enough qubits or channels to satisfy all the paths in a single time slot.

Computing the proposed routing metric EXT involves recursions, which may be slow for multi-path computation for all possible S-D pairs. Hence we propose three routing metrics, which are faster to compute. 1) \textbf{Sum of node distances (SumDist).} The success rate of a channel decreases exponentially with the physical distance $L$.

Hence SumDist can partially reflect the difficulty of a path.
2) \textbf{Creation Rate (CR).} CR is computed as $\Sigma 1/p_i$, where $p_i$ is the success rate of any channels on the $i$-th hop of the path. 3) \textbf{Bottleneck capacity (BotCap).} From \Cref{fig:E-hops-0.9,fig:E-hops-0.6}, the path width $W$ has greater impact on the path quality. The BotCap metric is $-W$, prefers wider path over narrower path, and uses the CR to order paths with the same width.

We consider the routing metric as a design parameter, and their efficiency is compared in \S~\ref{sec:eva}.

For each possible S-D pair, the server running the offline algorithm will use Yen's algorithm to get $L=25$ paths (\textit{offline paths}) for the pair, and tell \textit{each node in the network} about the resulting paths. And $L$ will grow by 50\% percent next time if the paths are not enough for the pair. An example is shown in Fig.~\ref{fig:QPass-algo}(a), the offline algorithm finds $ACDEB$ and $AC'D'DEB$ as two paths.

\subsubsection{P2 algorithm of Q-PASS}

The P2 algorithm runs on each node locally.
The inputs are all the offline paths $P$ (known before this time slot) and the S-D pairs received in P1 $O=\{o_i\}$, where $o_ i$ is an S-D pair $\langle s_i, d_i \rangle$.
The outputs are an ordered list of selected paths $P'$, each of which connects a single S-D pair. On the output, the local qubit-to-channel assignment and entanglement are performed on related nodes to build the links on these paths.
Since $P$ and $O$ are globally known for all nodes, the output $P'$ is also consistent on different nodes, similar to the global consistency of classical link-state routing.

The algorithm consists of two steps. 1) The paths computed from the offline algorithm for the current S-D pairs are retrieved and put into a priority queue, ordered by the routing metric. Then from the path with the lowest routing cost, the channels and qubits of the nodes on each path are reserved exclusively for each path. If a path has width $w$ by the offline algorithm, but currently available resource can only support width $0\leq w'<w$, then the path is reinserted to the queue with an updated metric calculated from $w'$. If $w'=0$, there is no available resource for the path, it is inserted to the back of the queue. This process ends until no paths can be fully satisfied.

2) After step 1, the queue contains all unsatisfiable paths in the ascending order of the routing metric. The qubits and channels for the satisfiable parts of each path (\textit{partial path}) are reserved in the order of the path in the queue. The partial paths can be used to recover link failures for the \textit{major paths} selected in step 1.
For the example of Fig.~\ref{fig:QPass-algo}(b), $ACDEB$ is reserved as the major path but $AC'D'DEB$ does not have enough resource. However, $AC'D'D$ can be reserved as a partial path.

When the two steps finish, each node assigns its qubits to the corresponding channels and try to generate quantum links. For example, $A$ in Fig.~\ref{fig:QPass-algo}(b) will assign one qubit to the channel to $C$ and another to the channel to $C'$.

\subsubsection{P4 algorithm of Q-PASS}

In P4, each node swaps the qubits locally to connect the links on the path, if the path is connected according to its local link states. However, link failures happen randomly, and the P4 algorithm focuses on the failure recovery based on the recovery paths established in P2. The inputs of P4 algorithm are: 1) S-D pairs from P1, 2) a major path list and a partial path list from P2, and 3) the $k$-hop link states of this node from P3.

We present two challenges of P4 algorithm: 1) the reserved paths may not always succeed because quantum links are probabilistic, and 2) after $k$-hop link state sharing in P3, no classical communication is allowed between any pair of nodes because the entanglement links decay quickly.

We propose segment-based path algorithm. The major path list is traversed from beginning to end. For each visited major path $\langle (v_0, v_1, \cdots, v_h), W\rangle$, it is divided into $\lceil h/(k+1)\rceil$ segments, each with width $W$: $(v_0, v_1, \cdots, v_{k+1})$, $(v_{k+1}, v_{k+2}, \cdots, v_{2k+2})$, $\cdots$, $(v_{\lceil h/(k+1) - 1\rceil (k+1)}, \cdots, v_{h-1}, v_h)$, such that each node knows the whole link states on the segment(s) containing it. Then for each `unresolved' failed link, recovery paths are found from the recovery path list, the mark the recovery path as resolved.

An example is shown in Fig.~\ref{fig:QPass-algo}. Assume $k=1$, and thus each node knows the link states of its 1-hop neighbors. The major path $ACDEB$ is divided into two segments $ACD$ and $DEB$, such that all nodes on a single segment know this segment is successful or not. If not, they will try to use a recovery path. In this example, $A$, $C$, and $D$ know link $C$-$D$ fails. Hence, the recovery path $AC'D'D$ is taken by both $A$ and $D$. The distributed recovery path selection is consistent because all recovery paths can be ordered deterministically.
Recall that a local policy is used to achieve distributed consistency for each path: each node places an intern link between the link with the smallest ID to its predecessor and the link with the smallest ID to its successor. And it repeats this process until no intern link can be made for this path.

\begin{figure}
	\centering
	\includegraphics[width=\linewidth]{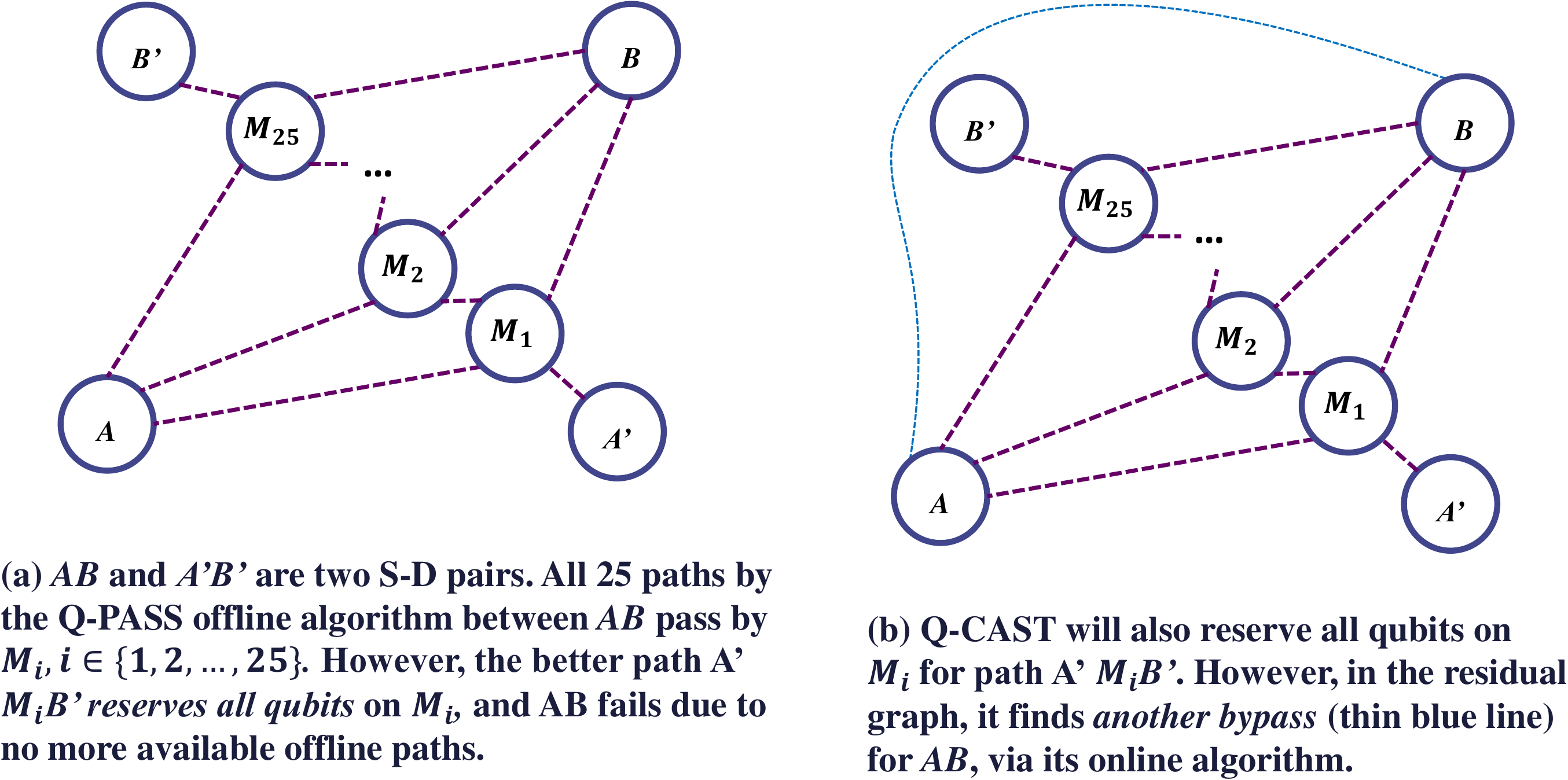}

	\caption{Comparison of Q-PASS and Q-CAST}
	\label{fig:QCAST-algo}

\end{figure}

\subsection{Q-CAST: Contention-free pAth Selection at runTime}
\label{sec:alg2}

The offline algorithm in Q-PASS has two fundamental disadvantages. 1) Since it does not know the actual S-D pairs when the paths are used, it has to compute the paths for all $n(n-1)/2$ pairs. 2) Besides the high computation cost, one more significant problem is that the computed paths may have severe resource contention. In Fig.~\ref{fig:QCAST-algo}(a), $AB$ and $A'B'$ are two S-D pairs. The offline algorithm of Q-PASS finds 25 paths for $AB$, all passing by any of the ${M_1, \cdots, M_25}$, then a single path for $A'B'$ may take all available qubits on $M_i$, and all 25 offline paths of $AB$ fail to be reserved in the residual graph, even though channels and qubits are available outside the offline paths. Due to unpredictable combinations of S-D pairs and hence the unpredictable residual graph after previously reserved paths, it is hard to pre-calculate paths for all S-D pair combinations.

Q-CAST does not require any offline computation and always finds the paths according to the topology and current reservations on qubits and channels.

\begin{figure}
	\centering
	\includegraphics[width=\linewidth]{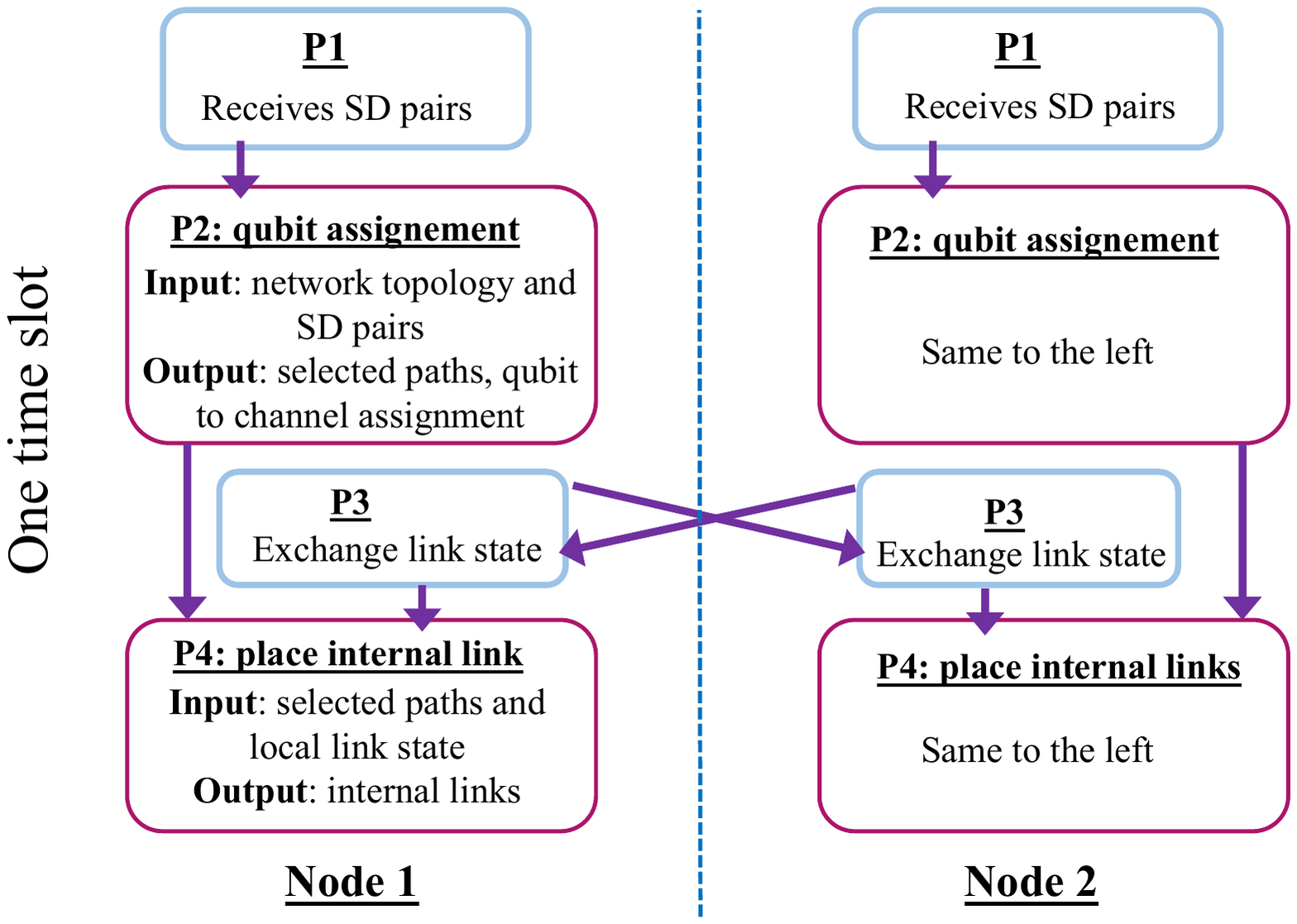}

	\caption{The algorithm workflow of Q-CAST}
	\label{fig:QCAST}

\end{figure}

\subsubsection{Algorithm overview}

The workflow of Q-CAST is shown in Fig.~\ref{fig:QCAST}. Q-CAST is a 4-phase algorithm, which does not require any offline computation. Q-CAST P1 and P3 are standard procedures similar to those of Q-PASS. Q-CAST P2 takes the input of the network topology and the S-D pairs. It finds and reserves paths one by one, \emph{without resource contention}. Recovery paths are also selected deterministically in P2. P4 takes the outputs of P2 and the link states from P3 to compute the swapping decisions.

\subsubsection{P2 Algorithm of Q-CAST}

The core task for Q-CAST P2 is to find multiple paths based on the knowledge of S-D pairs, and the paths are contention-free on qubits and channels. Yen's algorithm \cite{kshort} does not satisfy the requirement because its output paths are highly overlapped. Note, Q-PASS uses Yen's algorithm to find offline paths because the resulting overlapped path naturally provides small detours as recovery paths for major paths.
We propose to search multiple contention-free paths for online S-D pairs using a greedy algorithm, which runs as follows. Step 1) For every S-D pair, it uses the EDA (described later) to compute an optimal path in terms of the routing metric EXT between this pair. Step 2) Among the optimal paths of all S-D pairs, it selects the path with the highest EXT and reserve the resource of this path. Then the residual graph is computed by removing the reserved resource. 1) and 2) are repeated with the residual graph until no more path can be found, or the number of paths exceeds 200 -- a value to bound the number of paths in order to avoid unnecessary computation. We call this algorithm as \textit{Greedy EDA (G-EDA)}.

The above process aims to maximize the network throughput but does not consider fairness among S-D pairs. We will discuss how to balance throughput and fairness in a later section, which could be a future research topic.

\textbf{The optimal routing metric.} To find the optimal path under the EXT metric in a quantum network, the classical Dijkstra's algorithm fails because it only finds the shortest path when the routing metric is additive. Here, `additive' means the \textit{sum} of the costs of all edges on the path is exactly the cost of the whole path.  Obviously the ETX $E_t$ computed by Equation \ref{eq:P} is not additive.

We propose the \textit{Extended Dijkstra's algorithm (EDA)} to find the highest-ETX path in EXT between any S-D pair. The  highest-ETX path gives the maximum evaluation value among all possible paths between the S-D pair, with respect to a path quality evaluation function $e$. The input of $e$ is a path $\langle p, W \rangle$.

Similar to the original Dijkstra algorithm, EDA also constructs an optimal spanning tree rooted at the source node $s$. At the beginning the \emph{visited set} only includes $s$. The evaluation value from $s$ to an unvisited node $x$ is set as 0 or the evaluation value $e()$ of the edge $(s,x)$  if $s$ and $x$ are neighbors. Each time, the node $y$ with the maximum evaluation value to $s$ is added to the visited set and the evaluation values from $s$ to any other node $x$ are updated if $x$ and $y$ are neighbors. The algorithm stops when the destination is visited.
Different from the original Dijkstra algorithm, updating each evaluation value may cause the re-calculation of the evaluation function of the entire path, rather than simply adding the cost of a link. Though the updating may be complex, one optimization can be applied. If the path $p$ with hopcount $h$ and width $W$ grows by one hop and the new edge has at least width $W$, the width of the new path $p'$ stays unchanged to be $W$. The algorithm needs to calculate $E_t(p')=q^{h+1} \cdot \sum_{i=1}^{W} i \cdot P_{h+1}^i$ and $P_{h+1}^i$ is a recursive function. The original value of $P_h^i$ in calculating $E_t(p)$ can still be re-used, which significantly reduce the complexity of such recursion.

We skip the proof of the correctness of EDA due to space limit. Its correctness rely on a fact that the evaluation function $e$ of a path $\langle p, W \rangle$ should \textit{monotonically decrease} when extending $p$ to a longer path $p'$ by adding another node at the end of $p$. Since we use $E_t$ as the evaluation function,
we explain the monotonicity of $E_t$ without a strict proof. As the $p$ grows, $W$ may stay unchanged or decrease because the new edge may be narrower than $W$. In addition, adding one more hop will increase the risk of failure of the path. Hence, adding one hop means no wider path width and more hops to be transmitted, none of which can increase $E_t$.

\textbf{Bound the path length.} We set the upper-bound threshold $h_m$ of the path hopcount to ensure bounded path searching in EDA. During the EDA, for any path with hopcount larger than $h_m$, the path is ignored because it is unlikely to be a good path. The value of $h_m$ can be determined at system initialization. For a new network $G$, 100 pairs of nodes are randomly selected. Then, multipath routing is performed via G-EDA for each pair. The largest hopcount of the resulting paths whose $E_t>1$ is set to be $h_m$.

\begin{figure}
	\center
	\includegraphics[width=\linewidth]{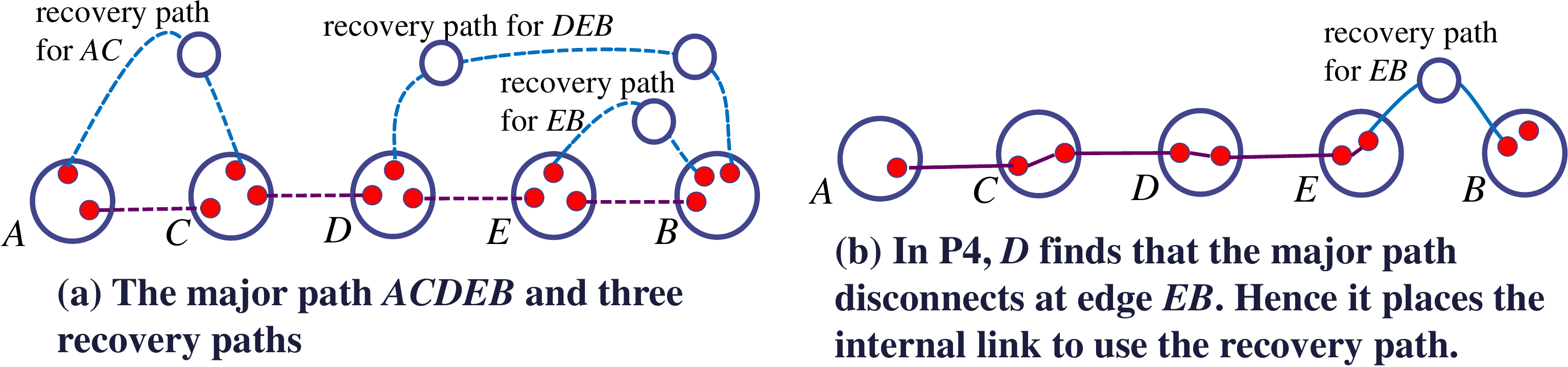}

	\caption{Example of recovery by Q-CAST}
	\label{fig:QCAST-recovery}

\end{figure}

\textbf{Recovery paths.} After finding the paths via G-EDA (denote as \textit{major paths}), the remaining qubits and channels can be utilized to construct \textit{recovery paths}, each of which ends at two nodes (denote as \textit{switch nodes}) on a single major path.

The switch nodes should be no more than $k$ hops away on a major path, where $k$ is the link state range, because in P4 the two nodes should ensure

consistent swapping decisions.

The recovery paths are found as following. For every node $x$ on a major path, we use EDA to find $R$ recovery paths  between $x$ and $y$ in the residual graph, where $y$ is another node that is $1$ hop away on the major path and $R$ is a small constant parameter. When all nodes are processed, the algorithm will run further iterations for the recovery paths that covers $l$ hops on the major path, for $l=2$ to $k$.
In Fig.~\ref{fig:QCAST-recovery}(a), the major path is $ACDEB$ and three recovery paths are found.

Every node will assign its qubits based on the reserved major paths and recovery paths, without any qubit/channel contention.

\subsubsection{P4 Algorithm of Q-CAST}

In P4, each node knows the major paths, the recovery paths, and the $k$-hop link states. It then makes the swapping decisions locally. The challenges for Q-PASS P4 still present for Q-CAST P4: probabilistic link failures and no interactive communication between nodes is allowed.

We propose an exclusive-or (\textit{xor, $\bigoplus$}) based algorithm to recover from potential link failures. We define the xor operator of two set of channels $C_1, C_2$: $C_1 \bigoplus C_2 = C_1 \cup C_2 \setminus (C_1 \cap C_2 )$. As both ends (switch nodes) of a recovery path $p_r$ are on a single major path, we can always find a sequence of nodes between the two switch nodes on the major path. This node sequence together with $p_r$ form a loop in the network graph, called a `\textit{recovery loop}'. Then, the link recovery algorithm works as following. The major path list is traversed from beginning to end. Each visited major path $\langle (v_0, v_1, \cdots, v_h), W\rangle$ is treated as $W$ separated 1-paths. For each 1-path, nodes find a set of recovery paths for failed links, such that the xor result of the major path and the recovery loops contains no hops on the failed links. To break the tie, shorter recovery paths are preferred. For example, in  Fig.~\ref{fig:QCAST-recovery}(a), $D$ and $E$ both find that the major path disconnects at edge $EB$. Also, both of them know that there are two recovery paths that can cover $EB$, namely the one from $D$ to $B$ and the one from $E$ to $B$. Hence the shorter one from $E$ to $B$ will be used, because shorter paths are likely to succeed than the long ones. $D$ still swaps qubits on the major path and $E$ switches to the recovery path.

The recovery algorithm is different from that of Q-PASS is because each recovery path in Q-CAST is dedicated to a single major path.

\section{Time and space costs}
\label{sec:analysis}

We denote the number of S-D pairs as $m$, and the maximum width of paths as $W_m$, which is determined by node capacities and edge widths.
We denote the maximum number of paths as $K_m$ in EDA. The number of nodes is $n$. We summarize the results here and some details together with the pseudocode can be found in the Appendix.

\textbf{Cost of routing metric evaluation.}

The time cost to calculate $E_t$ for a ($h,W$)-path according to Equation \ref{eq:P} is $O(hW)$, and the space cost is $O(W)$.

\textbf{Cost of P2 algorithm of Q-PASS.}

The time cost is $O(m K_m(h_m+\log(m K_m)))$ and the space cost is {\small$O(m K_m h_m+n)$}.

\textbf{Cost of EDA.} The time cost for EDA is $O(n\log n + |E|(h_m W_m))$. The space cost is $O(n)$.

\section{Performance Evaluation}
\label{sec:eva}

\subsection{Simulator Implementation}
\label{sec:impl}

\begin{figure}[t]
	\centering
	\begin{tabular}{p{110pt}p{110pt}}
		\centering\includegraphics[width=0.9\linewidth]{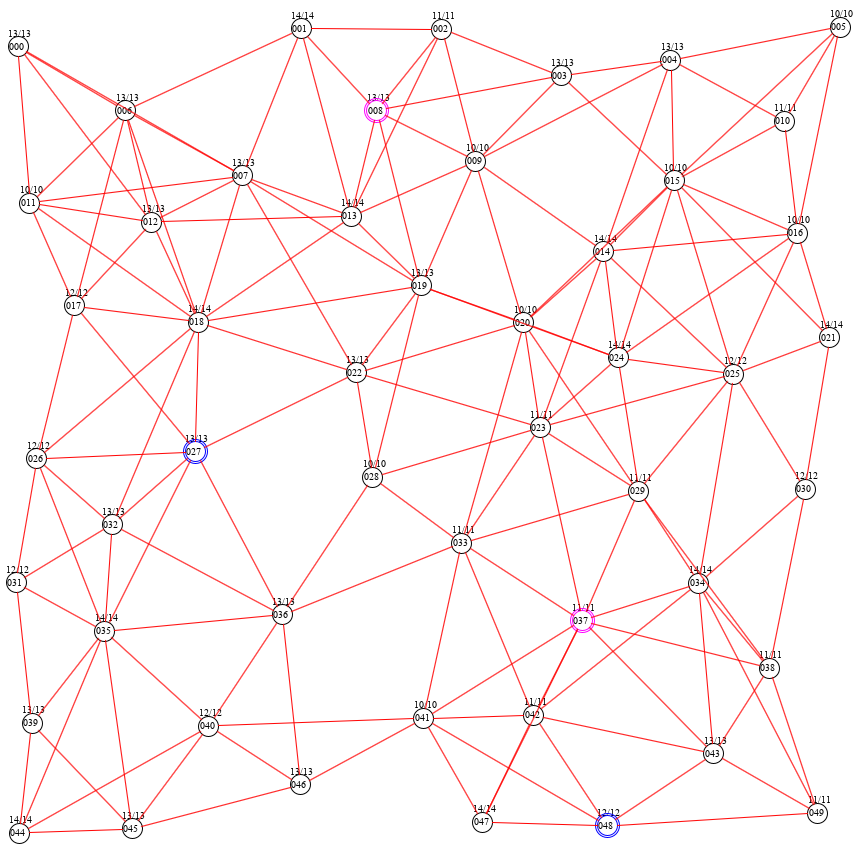}

		\caption{\footnotesize Visualized network with qubits and channels shown}
		\label{fig:visualizer-0}
		&
		\centering\includegraphics[width=0.9\linewidth]{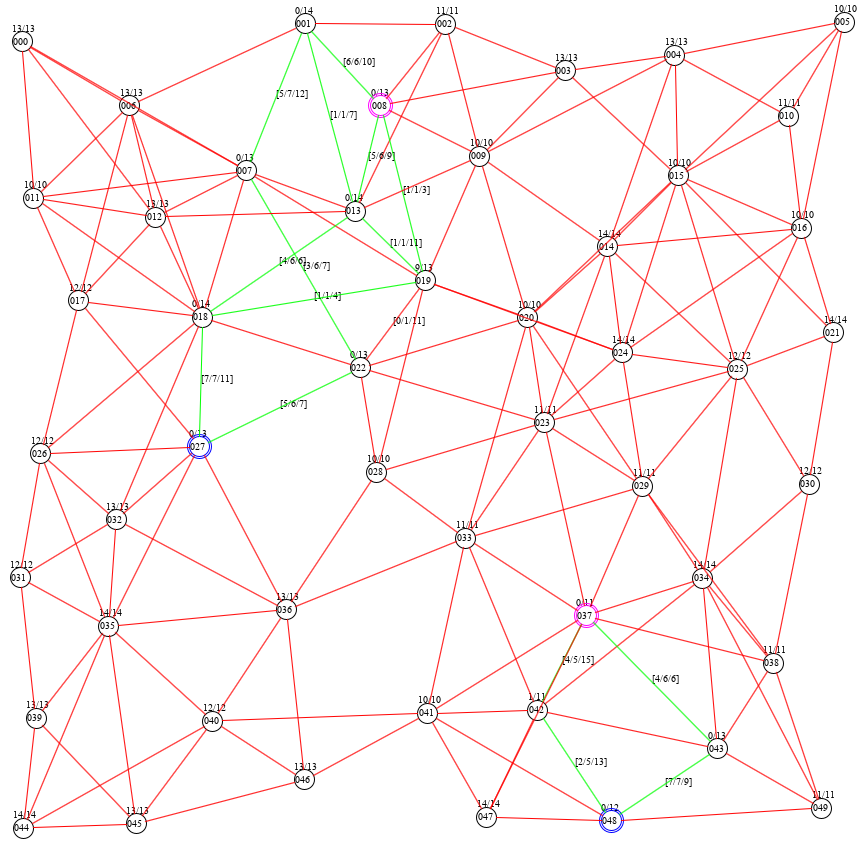}

		\caption{\footnotesize Visualized path selection and resource occupation}
		\label{fig:visualizer}
	\end{tabular}

\end{figure}

We implement the proposed network models and algorithms on a custom-built time-based simulator, with additional supports for topology generation, statistics, and network visualization. We do not use packet-based simulation because quantum networks do not use packet switching.

As shown in \Cref{fig:visualizer-0,fig:visualizer}, the visualization tool shows the network topology, current qubit/channel occupation, and existing quantum links at simulation runtime, for protocol analysis and demonstration.

The repository of the simulator is found on the anonymous link \cite{QuantumCode} and will be open to public for researchers to work on quantum network research.

We do not assume any specific topology and randomly generate quantum networks for simulations. We set the area $A$ holding quantum networks is a 100K units by 100K units square, each unit may be considered as 1KM. As the routing performance relies on the channel success rates rather than the length, we are not losing any generosity here. The network generation algorithm requires three input parameters: the number of nodes $n$, the average number of neighbors $E_d$, and the average success rate of all channels $E_p$. Nodes are randomly placed and the distance of any two nodes is at least $\leqslant 50/\sqrt{n}$ units. The edges are generated following the Waxman model \cite{Waxman} that has been used for Internet topology generators \cite{brite}.

After the topology generation, a binary search on the model parameter $\alpha$ is further carried out to make the average channel success rate to be $E_p \pm 0.01$. The number of qubits $Q$ for each node is independently uniformly picked from 10 to 14. The edge width $W$ is independently uniformly generated from 3 to 7, for each edge. We pick the range for $Q$ and $W$ based on our conjecture of a well-functioning quantum network. Our designs should work on wider ranges, which we cannot cover due to enormous possibilities.

\begin{figure*}[t]
	\centering
	\begin{tabular}{p{116pt}p{116pt}p{116pt}p{116pt}}
		\centering\includegraphics[width=\linewidth]{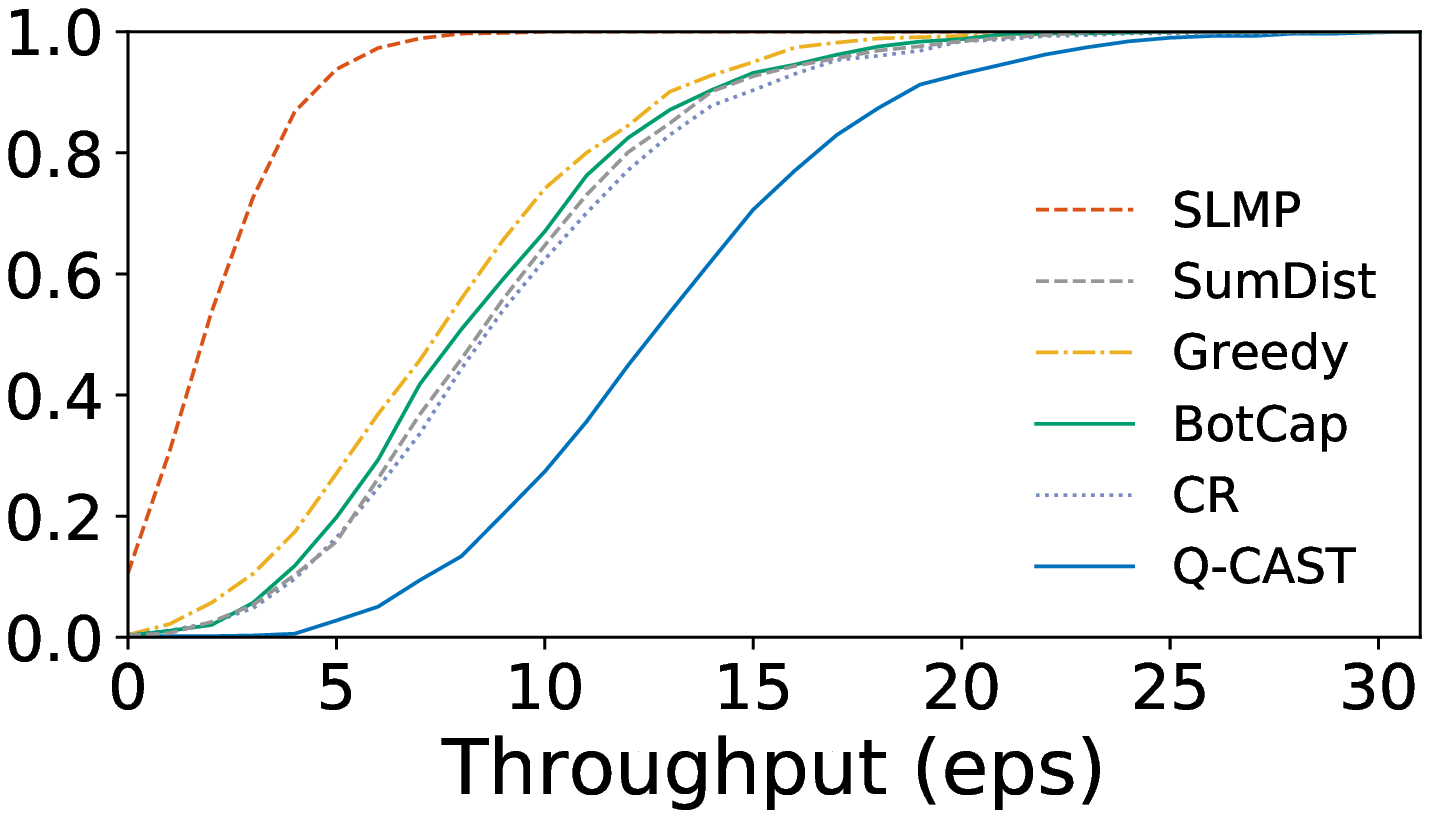}

		\caption{\footnotesize CDF of throughput under the reference setting}
		\label{fig:throughput-cdf}
		&
		\centering\includegraphics[width=\linewidth]{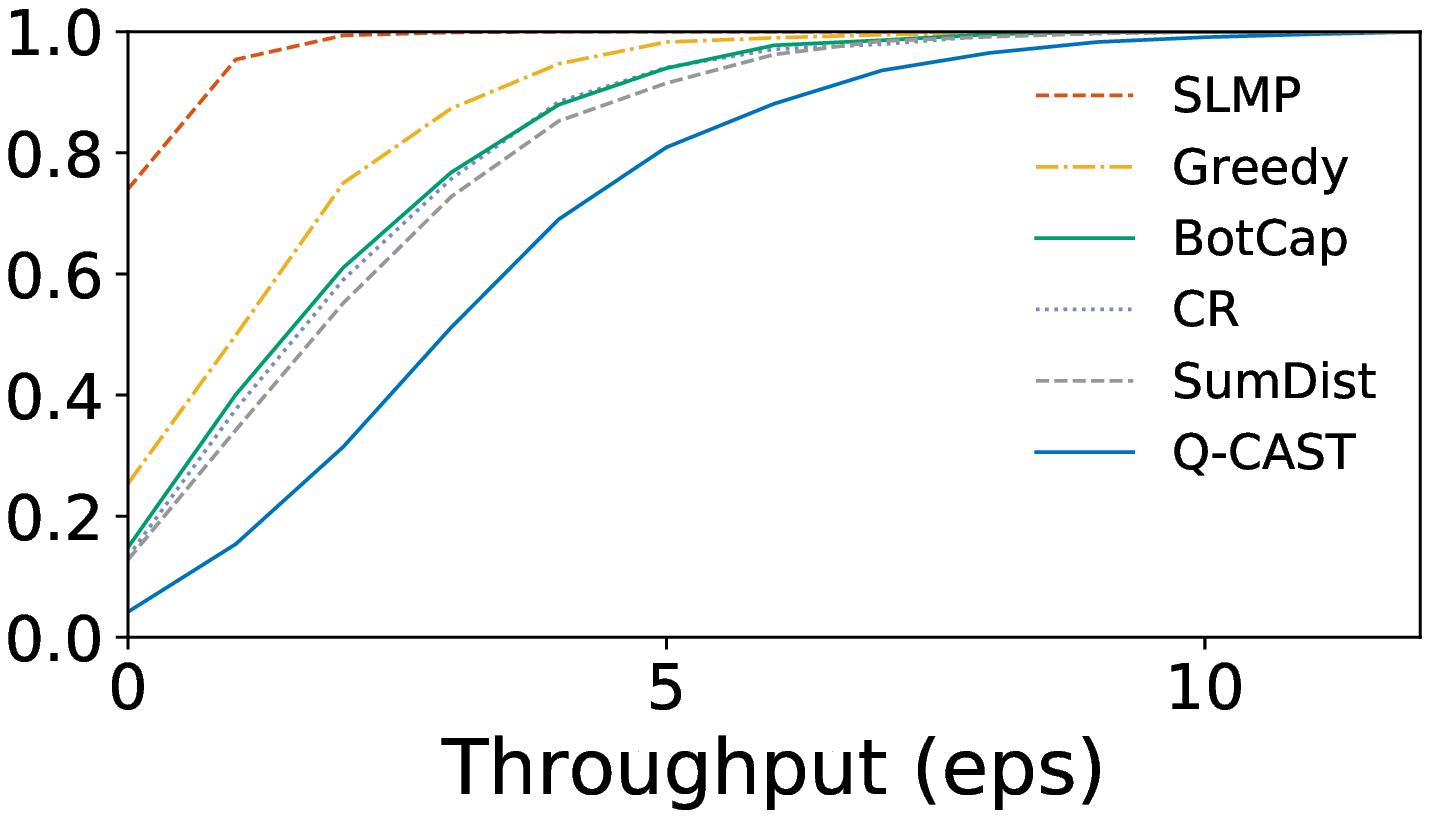}

		\caption{\footnotesize CDF of throughput, $E_p=0.3$}
		\label{fig:throughput-cdf-2}
		&
		\centering\includegraphics[width=\linewidth]{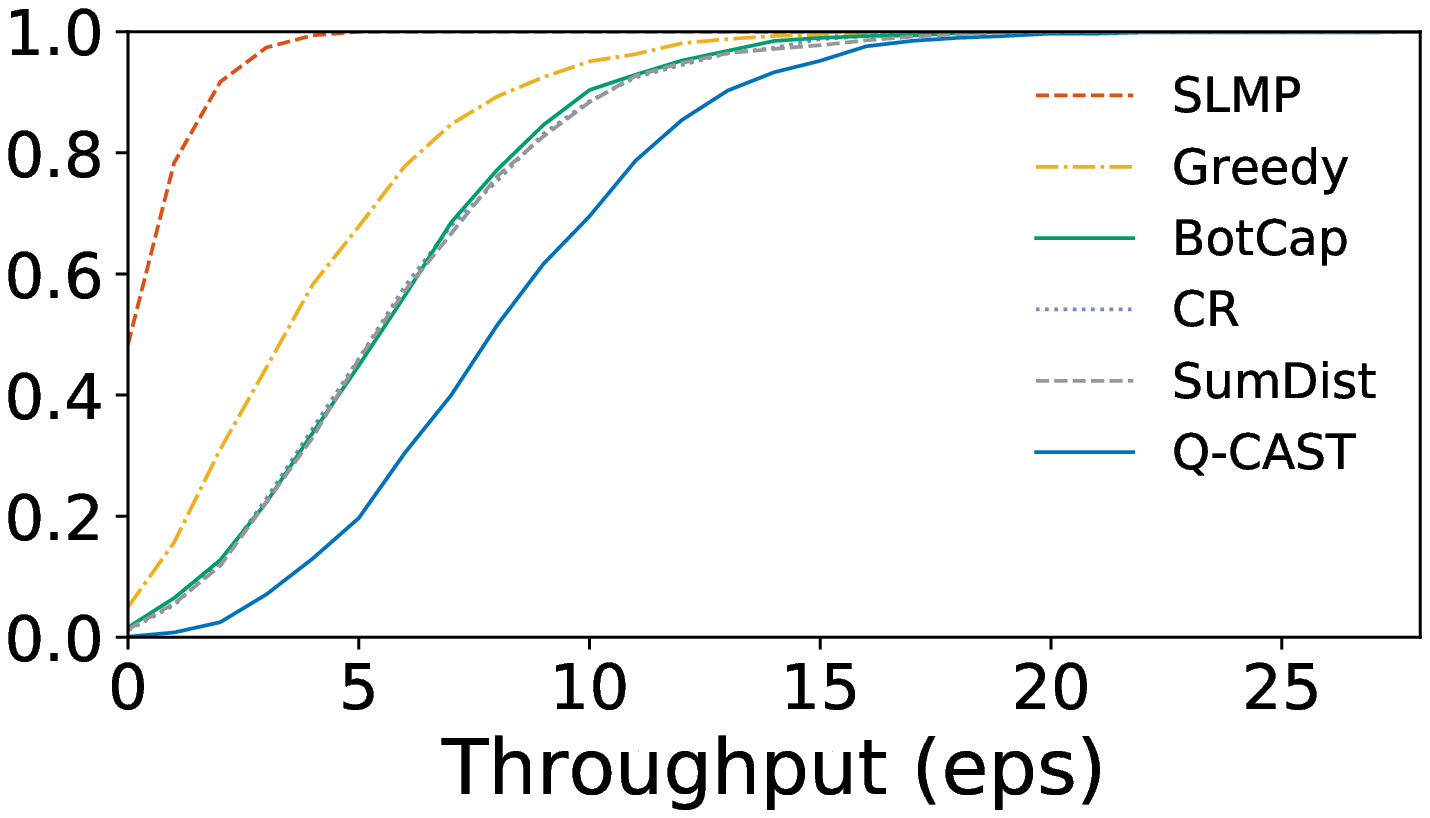}

		\caption{\footnotesize CDF of throughput, $n=400$}
		\label{fig:throughput-cdf-3}
		&
		\centering\includegraphics[width=\linewidth]{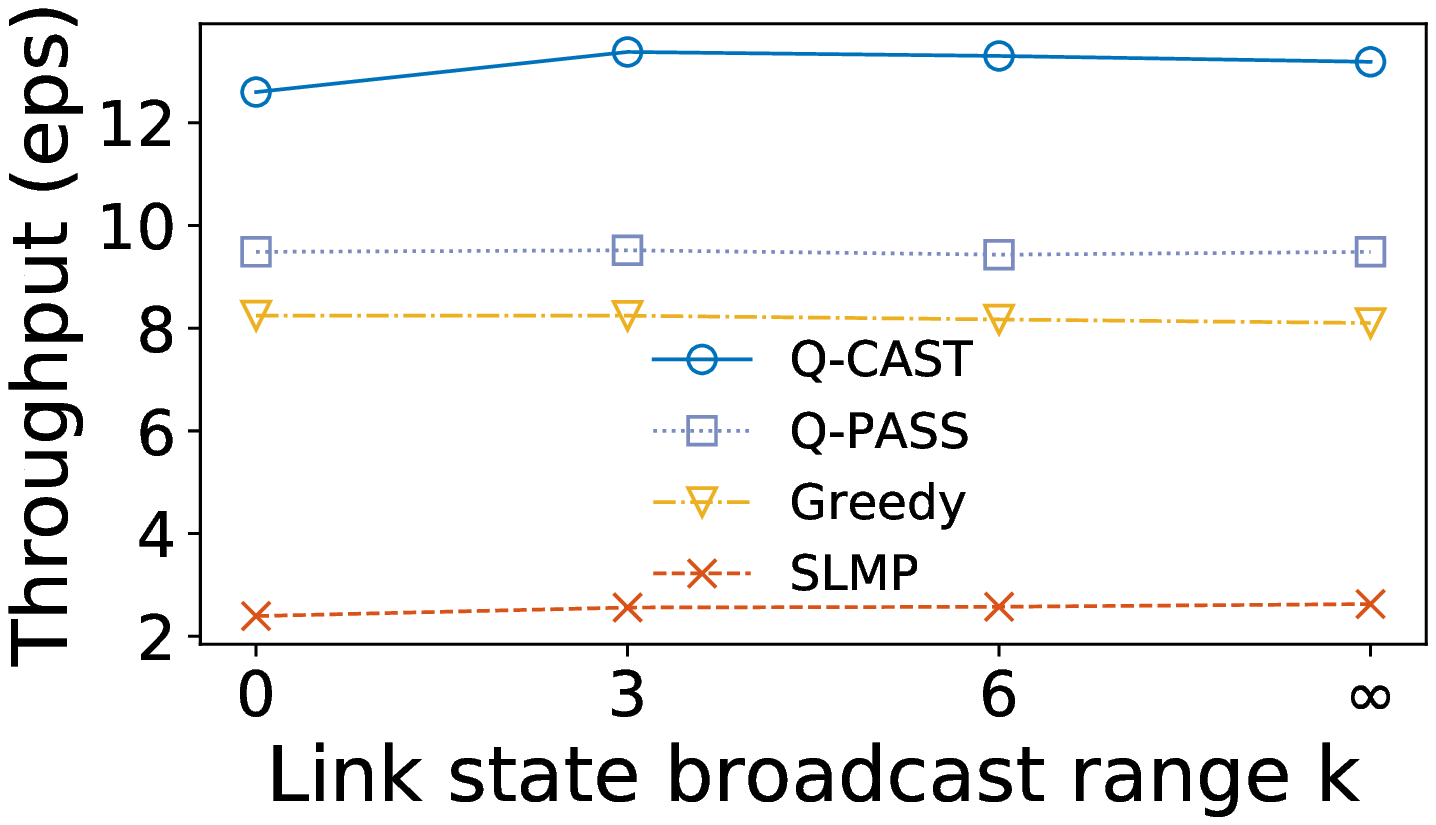}

		\caption{\small Throughput for different LS sharing ranges}
		\label{fig:throughput-k}
	\end{tabular}

\end{figure*}

\subsection{Methodology}

We evaluate the throughput, scalability, and fairness of the proposed entanglement routing algorithms. To gain the insight of the performance metrics and to provide reference for future research, we show more simulation statistics: the resource efficiency towards high throughput, the contribution of recovery paths for both algorithms.
Each data shown in the section is the average from 10 different network topologies.

We let the number of nodes $n$ vary in set $\{ 50, 100, 200, 400, 800\}$, average channel success rate $E_p$ vary in $\{0.6, 0.3, 0.1\}$, internal link success rate $q$ vary in $\{0.8, 0.9, 1.0\}$, link state range $k$ vary in $\{0, 3, 6, \infty\}$, average degree $E_d$ vary in $\{3, 4, 6\}$, and the number of S-D pairs $m$ vary from 1 to 10. To control variable, we show the results under the \textbf{reference setting} $n=100, E_p=0.6, q=0.9, k=3, E_d=6, m=10$, unless explicitly changed to observe the data trend. For each setting of $(n, E_p, q, k, E_d, m)$, 10 random networks are generated, and we simulate 1000 independent time slots on each of the network.

We compare Q-PASS and Q-CAST with two existing routing algorithms that have been used in quantum network studies: single-link multipath routing (\textit{SLMP}) \cite{RoutingEntanglement} and greedy routing \cite{GreedyRoutingQuantum}. Note the studies of \cite{RoutingEntanglement} and \cite{GreedyRoutingQuantum} are limited to special topologies such as circular or grid networks. Our results are the first to evaluate these algorithms on generalized topologies.

\begin{figure*}[t]
	\centering
	\begin{tabular}{p{116pt}p{116pt}p{116pt}p{116pt}}
		\centering\includegraphics[width=\linewidth]{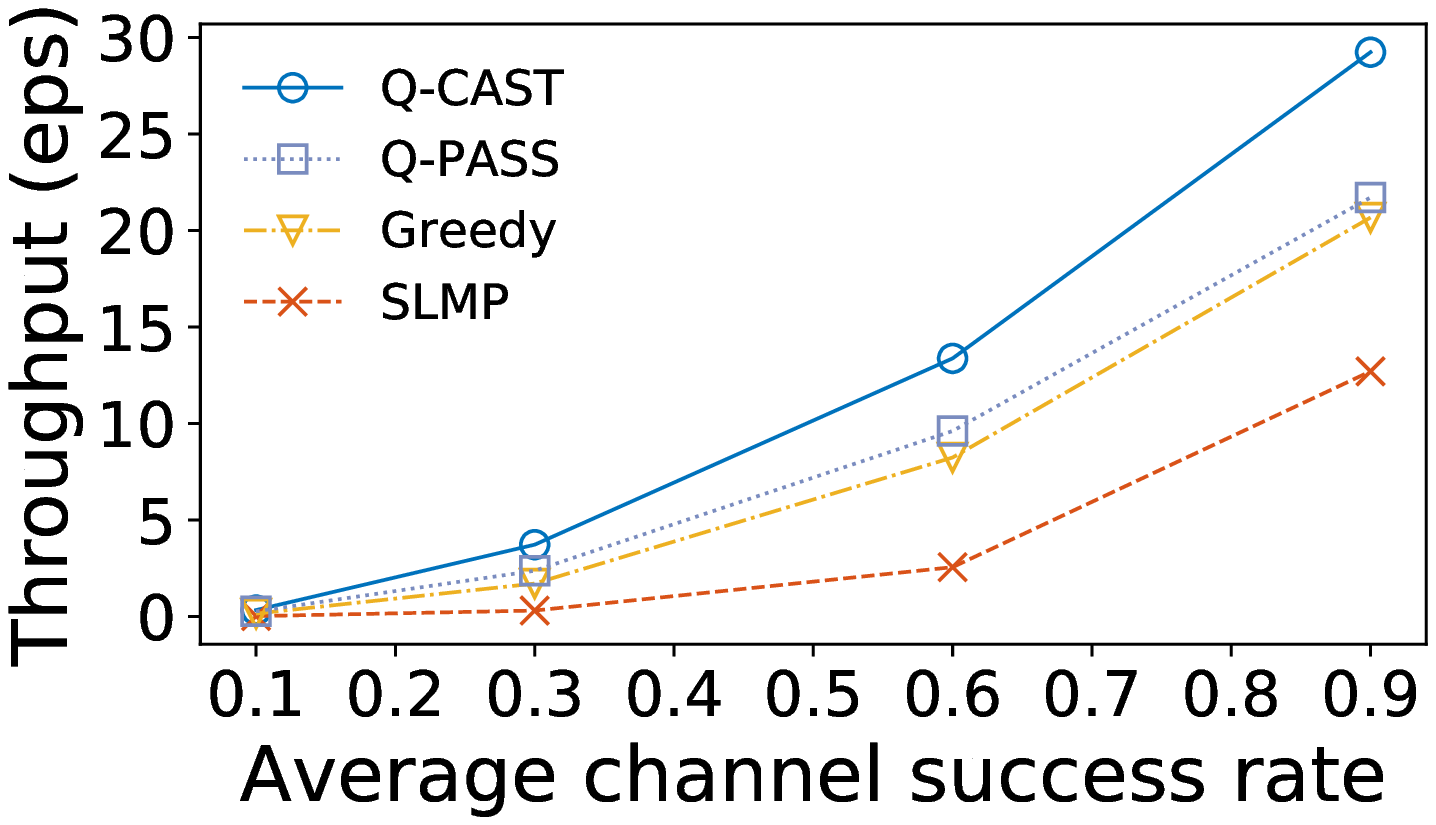}

		\caption{\small Throughput for different channel success rates}
		\label{fig:throughput-p}
		&
		\centering\includegraphics[width=\linewidth]{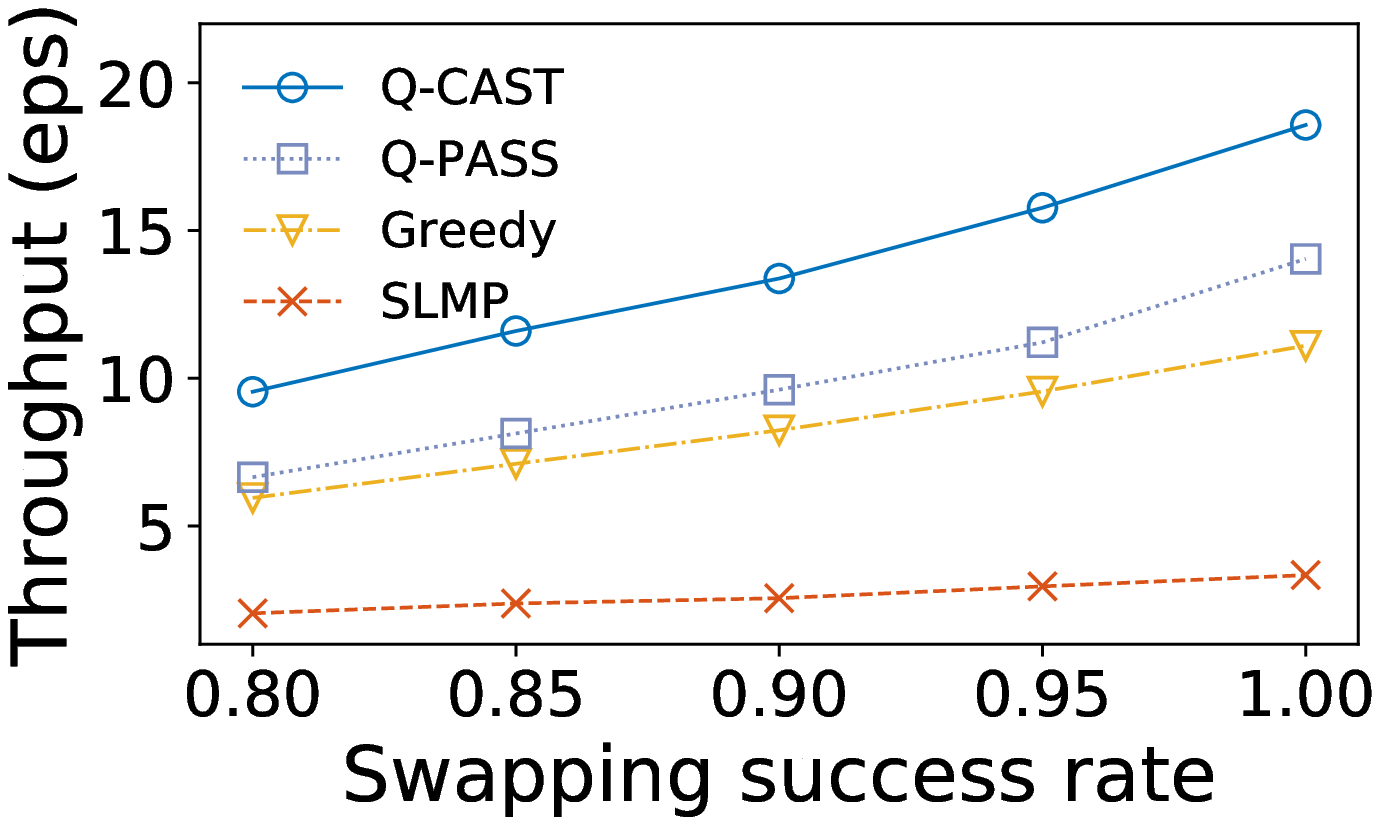}

		\caption{\small Throughput for different swapping success rates}
		\label{fig:throughput-q}
		&
		\centering\includegraphics[width=\linewidth]{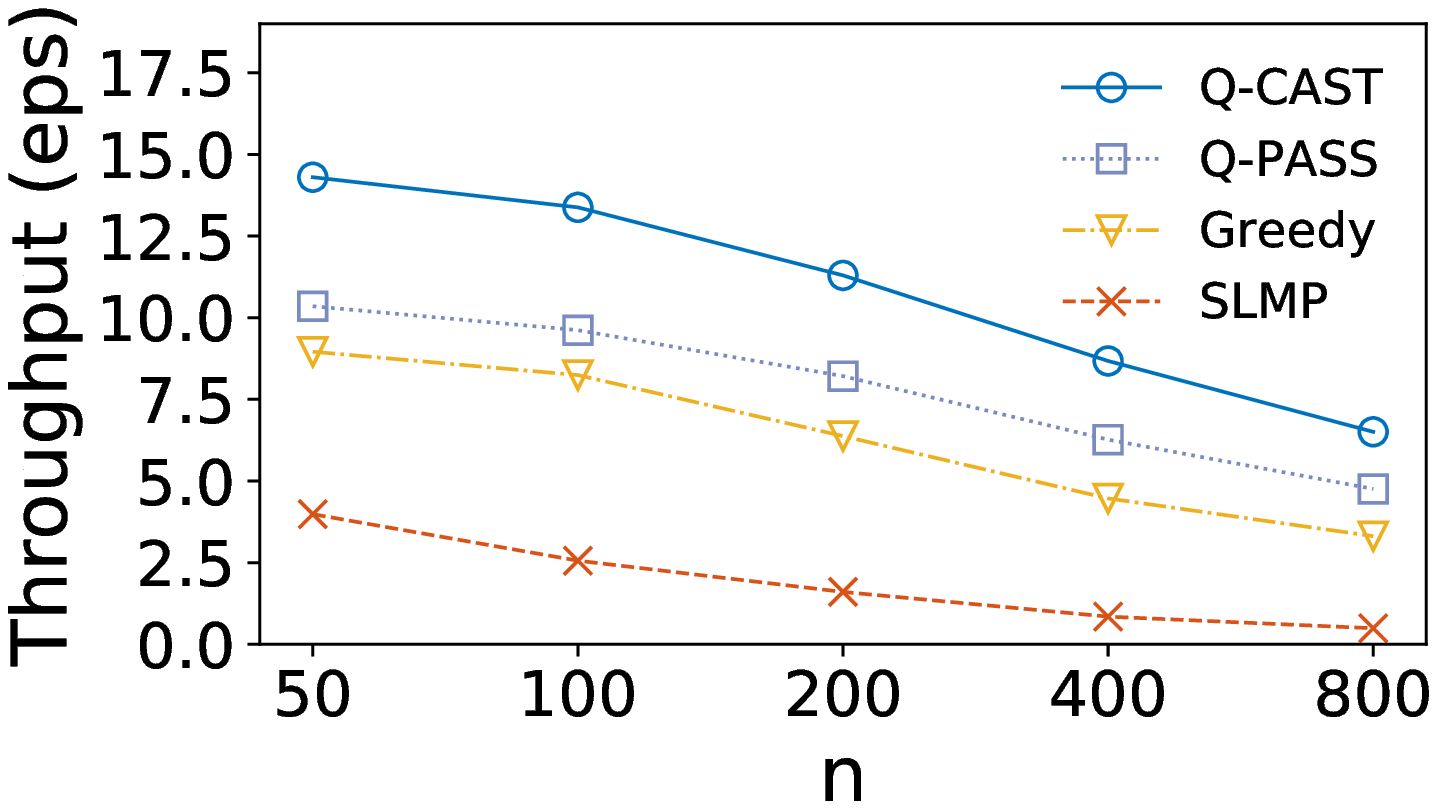}

		\caption{\small Throughput for different network sizes}
		\label{fig:throughput-n}
		&
		\centering\includegraphics[width=\linewidth]{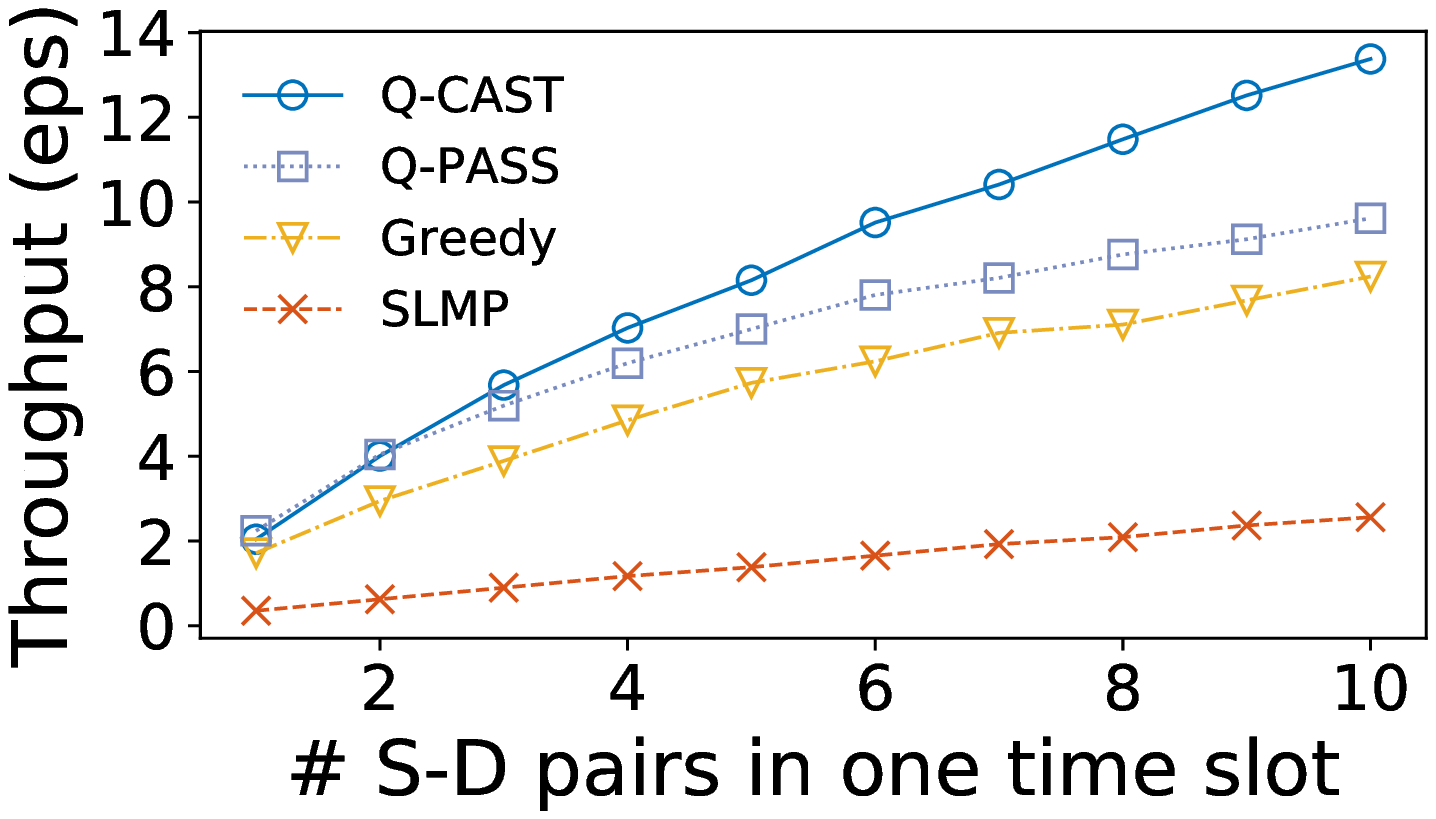}

		\caption{\small Throughput for different \# S-D pairs}
		\label{fig:throughput-nsd}
	\end{tabular}

\end{figure*}

\subsection{Evaluation results}

\textbf{Throughput. }
\Cref{fig:throughput-cdf,fig:throughput-cdf-2,fig:throughput-cdf-3} show the CDF of throughputs for Q-PASS, Q-CAST, Greedy, and SLMP, under the reference setting. The throughput results are calculated in terms of ebits per time slot (\textbf{eps}). The BotCap, CR, and SumDist are the routing metrics for the Q-PASS, and they are shown separately for better comparison. Despite the multipath routing, SLMP shows the lowest throughput because of the unreliability of a single channel/link. It fails to deliver any ebits in $>$10 percent of the time slots, and for 90 percent of the time slots, the total throughput between 10 S-D pairs are less than 5. The Greedy enjoys a high throughput, and for more than 90 percent of the time, it delivers more than 15 ebits for 10 random S-D pairs. For Q-PASS, all the three metrics of it exhibit similar throughput, and the the CR metric gives the highest throughput among all metrics, which delivers about 2 eps more than the Greedy. Q-CAST shows great advantages over all other algorithms and outperforms the CR about 5 eps. Q-CAST is also the most reliable because it seldom delivers less than 5 eps. Since CR is slightly better than other metrics, we use CR to represent Q-PASS in the following results.

\textbf{Vary link state range. }
In P3, each node shares its link states with its $k$-hop neighbors, and hence, $k$ influences the path recovery performance. Fig.~\ref{fig:throughput-k} shows the average throughput on different $k$. The Greedy algorithm does not rely on $k$ and is shown for reference. $k$ contributes little to the overall performance because most path failures are just one hop $v_i$-$v_{i+1}$, which can be recovered by $v_i$ and $v_{i+1}$ with their own link states. $k=3$ is sufficient for Q-CAST, and larger $k$ slightly degrades the throughput because longer and more unreliable recovery paths may be selected. This would occupy the routing resource which could have been allocated to other shorter and more reliable recovery paths.

\textbf{Vary link success rates. }
\Cref{fig:throughput-p,fig:throughput-q} show the average throughput of Q-PASS, Q-CAST, Greedy, and SLMP on different quantum device abilities by varying the average channel success rate and swapping (internal link) success rate. When the channel success rate $p$ or the swapping success rate $q$ is small, the overall throughput will be degraded. A robust routing algorithm should still perform well on low ability networks.
From the figures, the swapping success rate also has big impact on the average throughput, because the link failure in the P2 can be mitigated by the recovery algorithms in P4, but there is no circumvention for swapping errors. And the Q-CAST performs the best among the four algorithms.

\textbf{Scalability. }
We evaluate the scalability of routing algorithms on two dimensions: the size of the network $n$ and the number of concurrent S-D pairs $m$. A larger network means the average distance of S-D pairs is longer; and more concurrent S-D pairs in one time slot introduce higher level of resource contention. Fig.~\Cref{fig:throughput-n,fig:throughput-nsd} show the average throughput on the two dimensions. All algorithms exhibit a logarithmic throughput decrease with the number of nodes in the network. Q-CAST outperforms others on all network sizes, and the throughput of Q-CAST is as high as 7.5eps when the network contains 800 nodes.
The reason of lower throughput in larger networks is because the average path length is longer for the S-D pairs. Longer paths are more likely to fail in quantum networks.
Besides, the throughput of all algorithms grow sub-linearly with the number of S-D pairs, due to resource contentions. Q-CAST outperforms others on most settings, and the advantage of Q-CAST over other algorithms grows rapidly with the number of S-D pairs. It is because Q-CAST actively resolves the resource contentions for the S-D pairs.

\begin{figure*}[t!]
	\centering
	\begin{tabular}{p{116pt}p{116pt}p{116pt}p{116pt}}

		\centering\includegraphics[width=\linewidth]{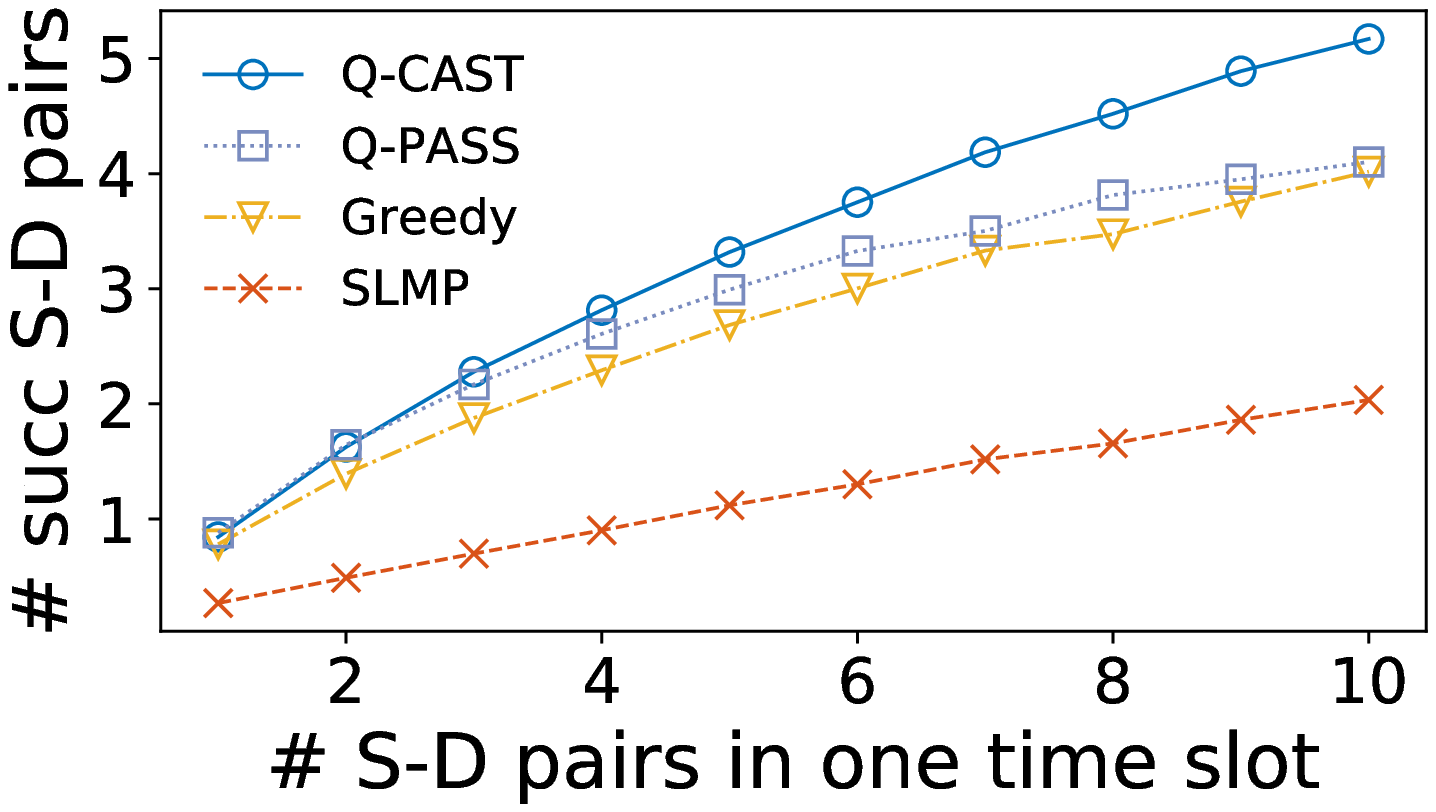}

		\caption{\small \# successful concurrent S-D pairs}
		\label{fig:succ-pairs-nsd}
		&
		\centering\includegraphics[width=\linewidth]{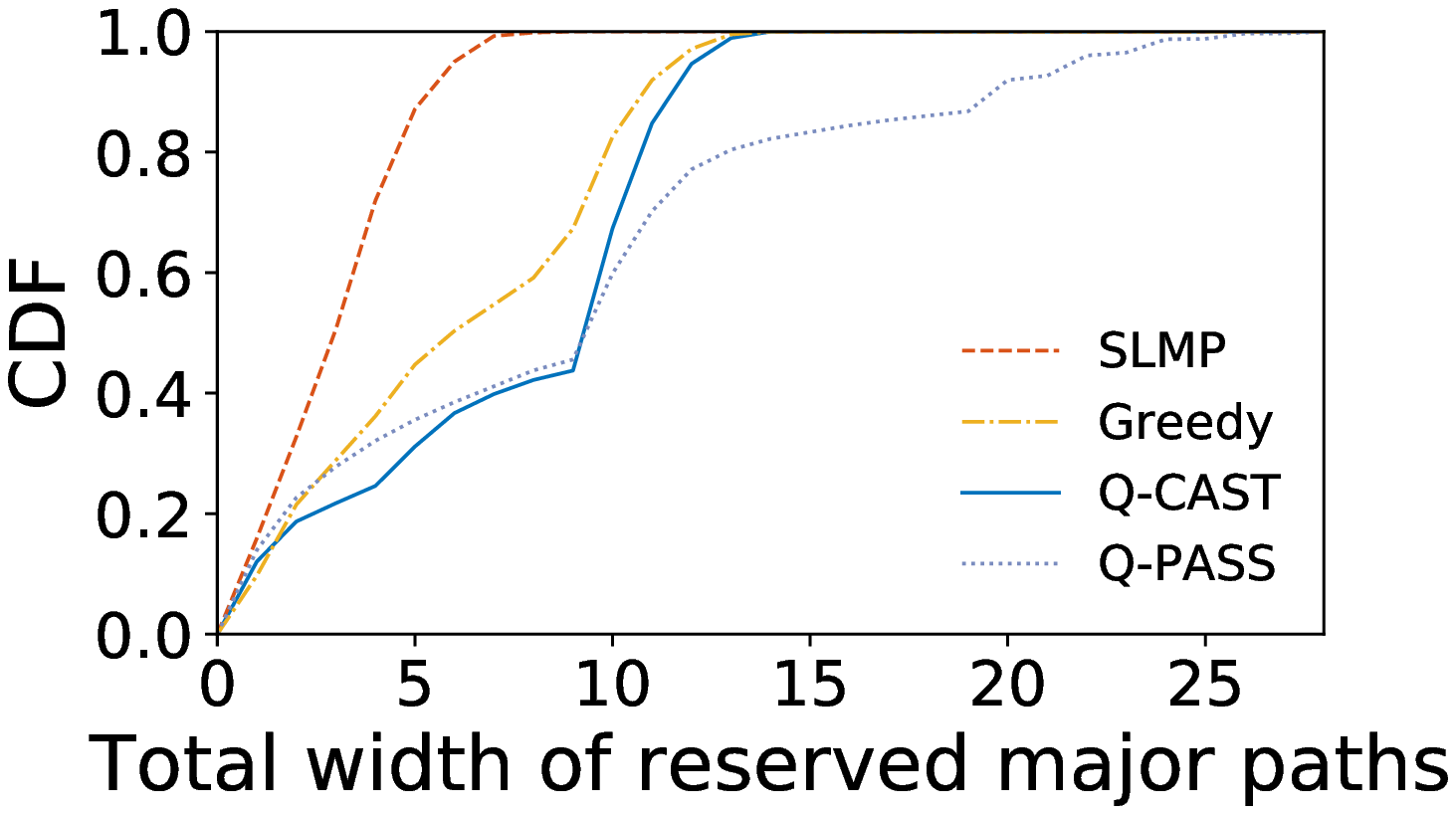}

		\caption{\footnotesize CDF of \# major paths under the reference setting}
		\label{fig:mp-width-cdf}
		&

		\centering\includegraphics[width=\linewidth]{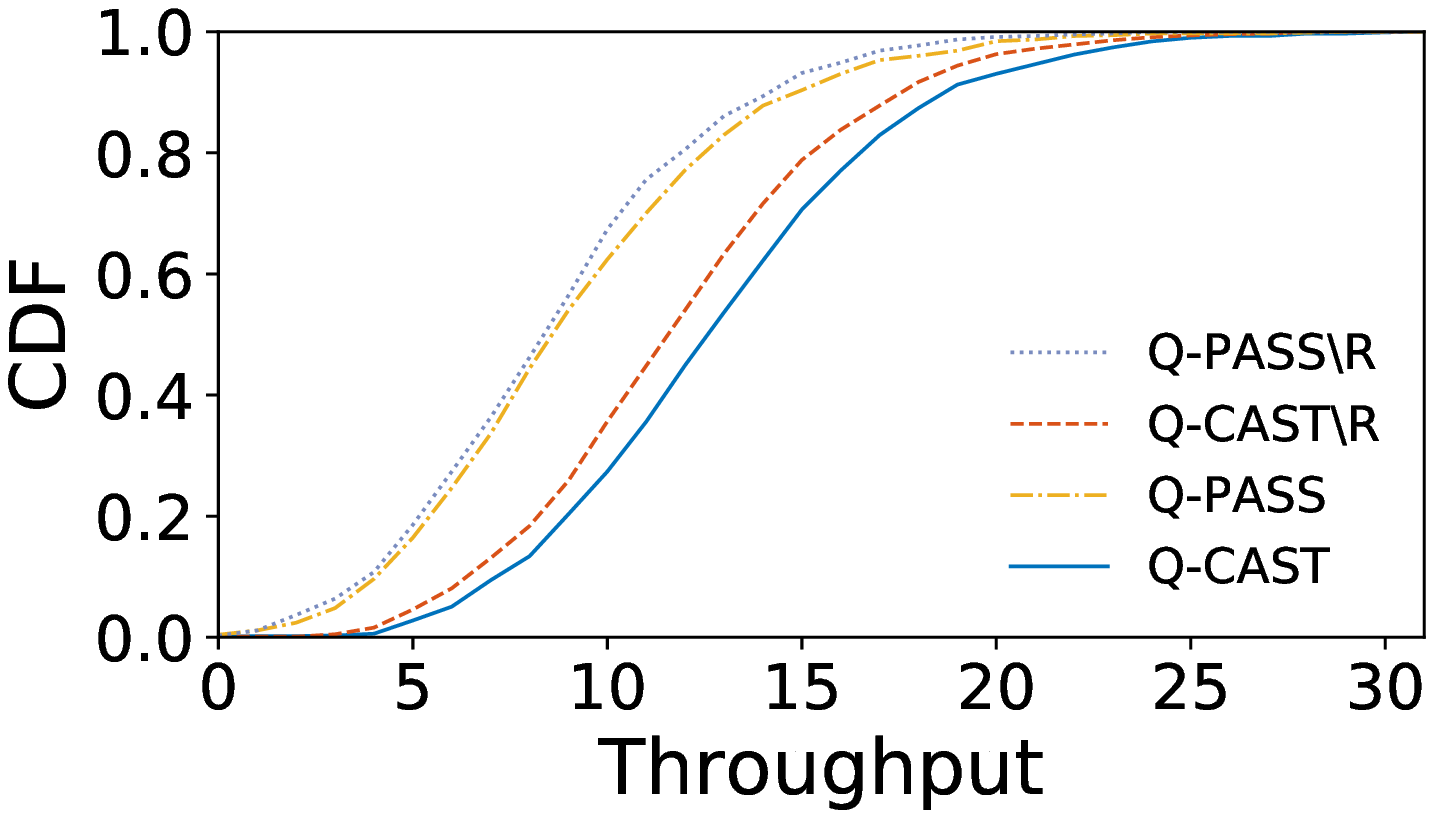}

		\caption{\small Contribution of recovery paths}
		\label{fig:rp-throughput-cdf}
		&
		\centering\includegraphics[width=\linewidth]{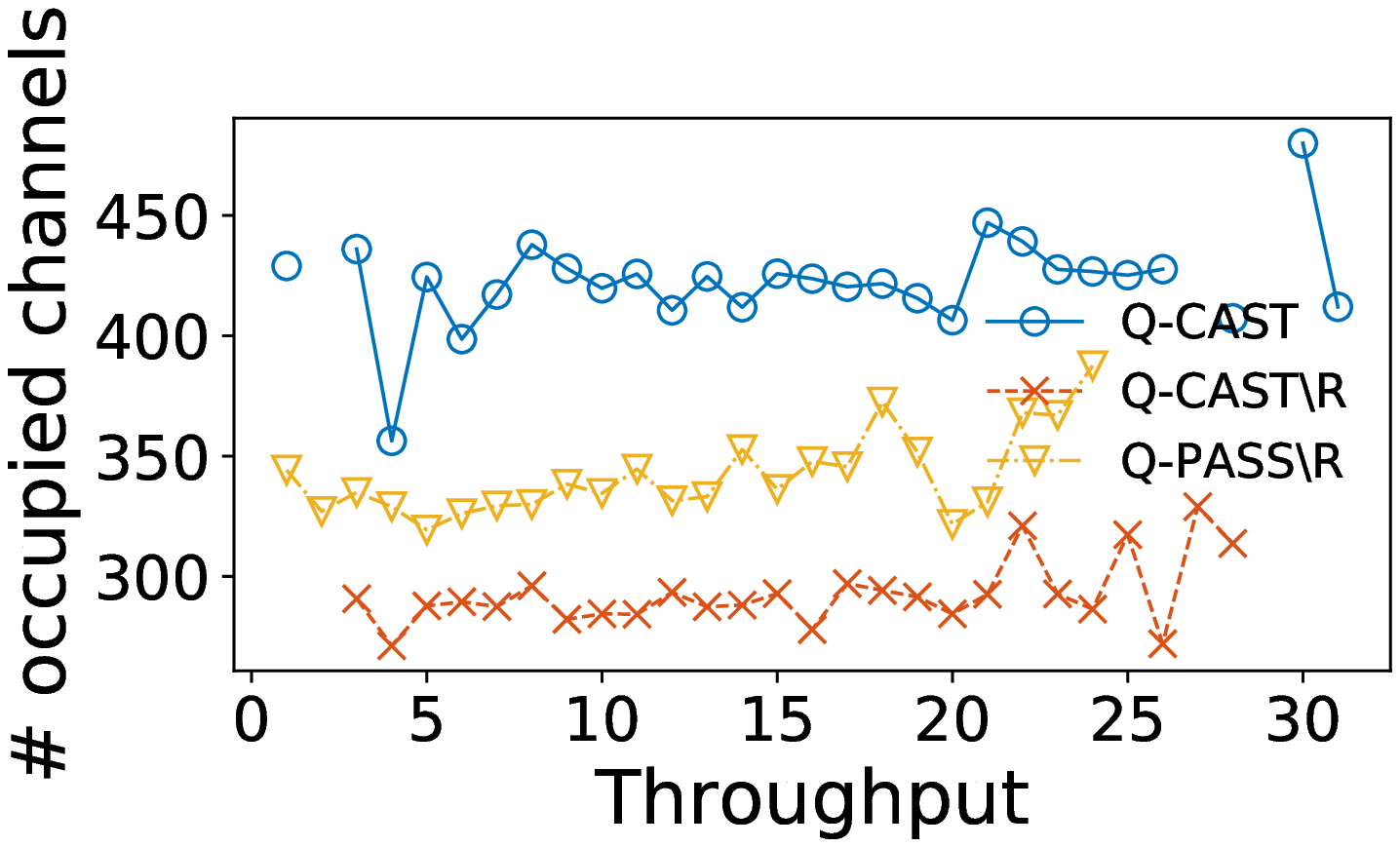}

		\caption{\small Overhead of recovery paths}
		\label{fig:channels-throughput-rp}
	\end{tabular}

\end{figure*}

\textbf{Fairness. }
Though we aim to maximize the throughput in the current designs, the fairness among the S-D pairs are evaluated. Fig.~\ref{fig:succ-pairs-nsd} shows the average number of successful S-D pairs under different number of concurrent requests. For a time slot, a S-D pair is successful (\textit{epair}) when they establish at least one ebits after P4.
Q-CAST outperforms others and all algorithms grows sub-linearly.
Fig.~\ref{fig:mp-width-cdf} shows the CDF of the number of paths allocated to every S-D pair. A $W$-path is counted as $W$ separate paths. As a base line requirement, any S-D pair should be allocated at least one major paths, which is fulfilled by all algorithms. The SLMP is the most fair. The Q-CAST has a turning point on the CDF figure, which means 40 percent of S-D pairs are allocated less than 9 paths, and the other pairs are allocated 10 to 14 paths, which is very fair. The Q-PASS is the most biased algorithm.

\textbf{Recovery paths. }
We evaluate the contribution of recovery paths to the overall throughput for both Q-PASS and Q-CAST, by comparing their throughput with that of their recovery path-free versions Q-PAST$\backslash$R and Q-CAST$\backslash$R. The results are shown in Fig.~\ref{fig:rp-throughput-cdf}. The recovery paths contribute about 0.5eps to Q-PASS and 1eps to Q-CAST. We further show the average number of occupied channels in one time slot for Q-CAST, Q-CAST$\backslash$R, and Q-PASS in Fig.~\ref{fig:rp-throughput-cdf}, where the $x$-axis show the throughput of each case. Q-PASS is not shown in this figure because it takes way more channels in the recovery paths and the results are not in this range of $y$-axis. Q-PASS$\backslash$R takes times less channels compared with Q-PASS, and Q-CAST$\backslash$R saves 25\% channels from the 400 channels taken by Q-CAST.

\begin{figure}[t]
	\centering
	\begin{tabular}{p{116pt}p{116pt}}
		\centering\includegraphics[width=\linewidth]{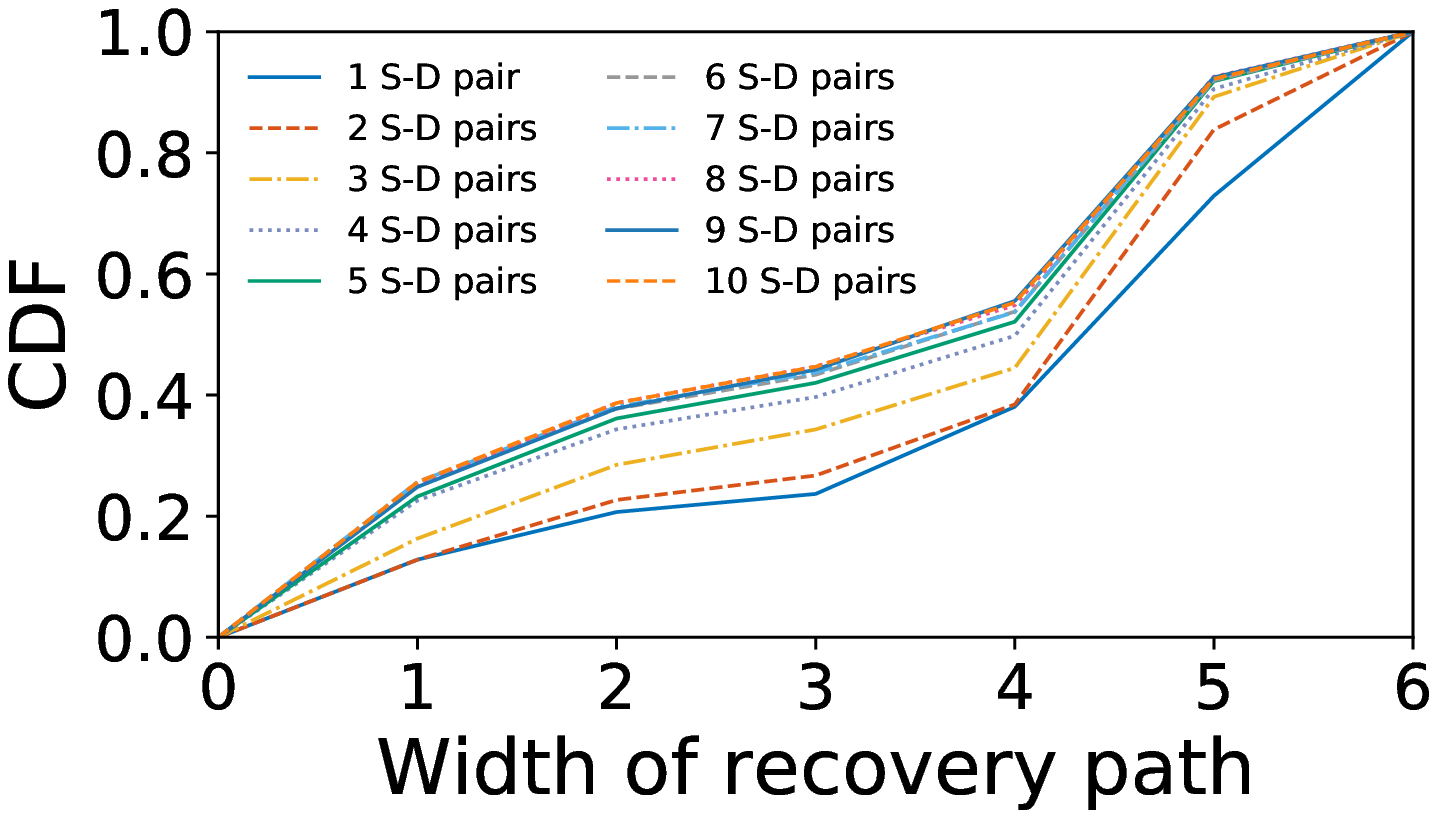}

		\caption{\small CDF of the width of recovery paths}
		\label{fig:a2-rp-wid-cdf-6-100-06-09-3-nsd}
		&
		\centering\includegraphics[width=\linewidth]{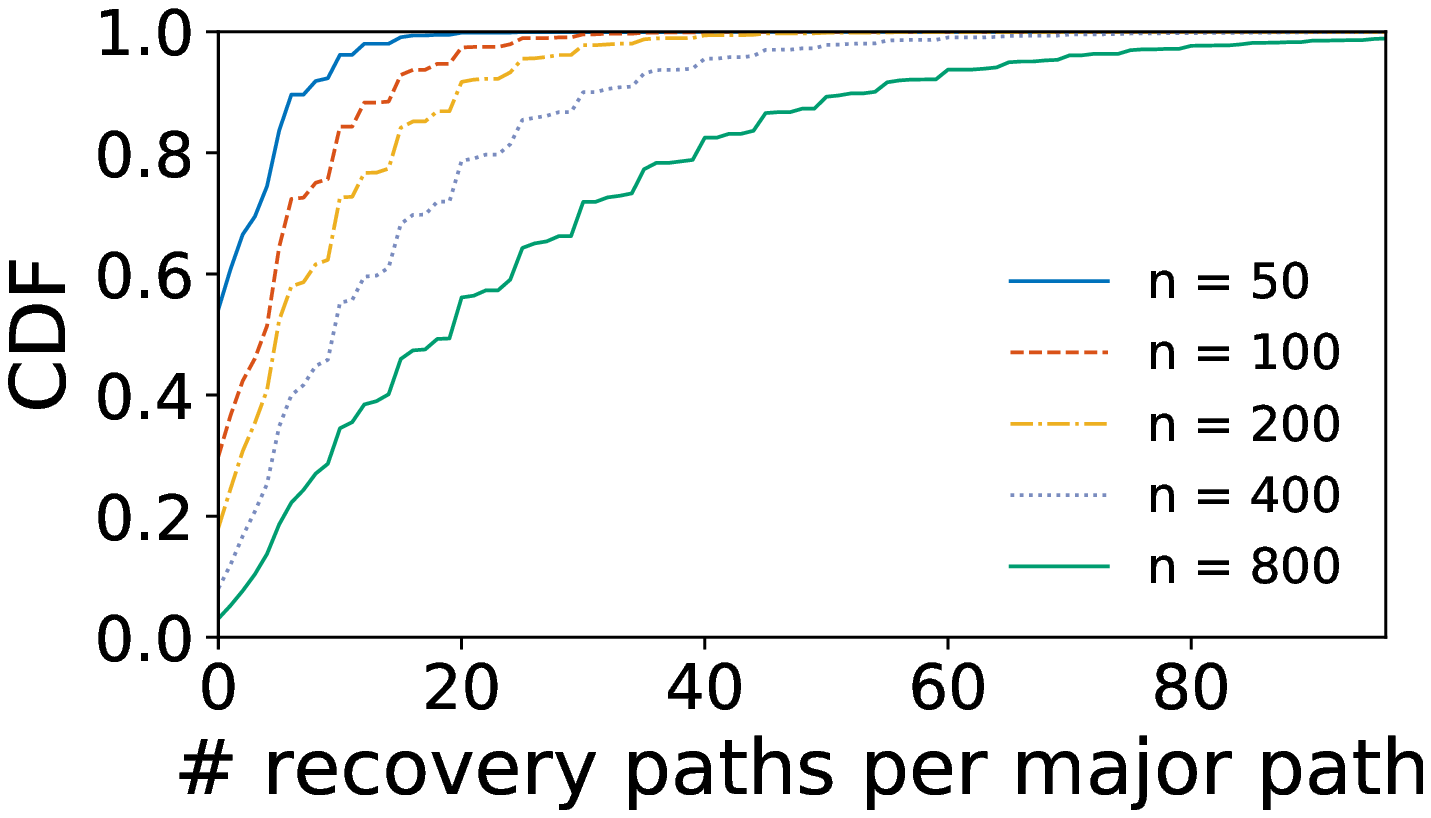}

		\caption{\footnotesize CDF of \# recovery paths on a single major path}
		\label{fig:a2-rp-wid-per-mp-cdf-6-n-06-09-3-10}
	\end{tabular}

\end{figure}

As the recovery paths are contention-free for Q-CAST, more interesting statistics are collected on Q-CAST recovery paths. The CDF of the width of recovery paths is shown in Fig.~\ref{fig:a2-rp-wid-cdf-6-100-06-09-3-nsd}. The recovery paths can be wider when the number of S-D pairs is small, because of the low resource contention between S-D pairs. For most cases, the widths of recovery paths for a single S-D pair is larger than those of the 10 concurrent S-D pairs by 2. Besides, the CDF of total number of recovery paths of a single major path is shown in Fig.~\ref{fig:a2-rp-wid-per-mp-cdf-6-n-06-09-3-10}. In larger networks, the major paths are longer and more recovery paths can be found.

\textbf{Summary of evaluations. }
Q-CAST exhibits much higher throughput, robustness, and scalability than other routing algorithms. SLMP has much lower overall throughput and Greedy's throughput is between SLMP and Q-CAST.

Q-PASS also shows good throughput and the metric CR provides the highest throughput for Q-PASS.
If the minimum resource utilization is a concern for some quantum networks, recovery paths for both algorithms can be disabled for best efficiency. Q-CAST$\backslash$R is a good balance between throughput and resource efficiency.

\section{Discussion}

\label{sec:discuss}

\textbf{Better fairness. } The algorithms proposed in this paper aim to maximize throughput, and each time slot is considered totally separately. A simple extension, however, is available to both Q-PASS and Q-CAST to provide better fairness while maintaining high throughput. For any S-D pair that has failed to share an ebit in a slot $T$, the pair and the failing streak $\langle (s, d), 1 \rangle$ are broadcast to all nodes in P1 in the slot $T+1$. The routing metric of all paths connecting this S-D pair is multiplied with a factor such as $1.1$, which means their paths are slightly over-evaluated, and thus are more likely to be selected. If the pair still fails, the failing streak increases to a higher factor such as  2, and the related routing metric is multiplied with $1.1^2$ in $T+2$. Eventually, this pair will succeed.

\textbf{Prioritized routing. } Both Q-PASS and Q-CAST are extendable to support simple prioritized routing. Suppose S-D pairs are in different priority classes, identified by the number $1, 2, \cdots, 10$, and the priority is `hard' -- a single S-D pair in priority class $c$ is far more valuable than all S-D pairs in priority class $c-1$ and lower. In P2 of Q-PASS and Q-CAST, the offline paths (only Q-PASS) and online paths (both algorithms) of the highest priority S-D pair are selected until no more path is available. More paths are then selected in the residual graph for S-D pairs in lower priority class. The P4 of Q-CAST is not modified because the selected paths have no contention. In P4 of Q-PASS, the paths of the highest priority S-D pair are recovered first.

\textbf{Continuous model. } 
The time slot model used by this work is called the on-demand model in \cite{GreedyRoutingQuantum,RoutingEntanglement}.
The continuous model is proposed in \cite{GreedyRoutingQuantum}, which assume a quantum link established in previous slots and possibly be used in later slots.  Under the continuous model, failed links can be retried throughout multiple time slots while holding the rest of the path, until every link on a path is successful. Then the path is built via quantum swapping. However, this approach only works for a single S-D pair. If concurrent S-D pairs exist, routing may easily fail because some links are occupied by the holding paths, unless there is global link state broadcast that causes high communication cost and long latency. 

We consider the continuous model as another direction which requires a co-design of the network topology and the routing algorithm.

\section{Conclusion}

\label{sec:conclusion}
This work presents a new entanglement routing model of quantum networks that reflects the
difference compared to  classical networks and new entanglement routing algorithms that utilize the unique
properties of quantum networks. The proposed algorithm Q-CAST increases the network throughput by a big margin compared to other methods. We expect more future research will be conducted to the entanglement routing problem and could contribute to the success of quantum networks.

\clearpage
{\small
	\bibliographystyle{plain}
	\bibliography{bibfile}
}

\appendix
\section{Appendix}

\subsection{Finding the optimal path selection for Q-CAST}

\begin{figure}[h]
	\centering
		\centering\includegraphics[width=.6\linewidth]{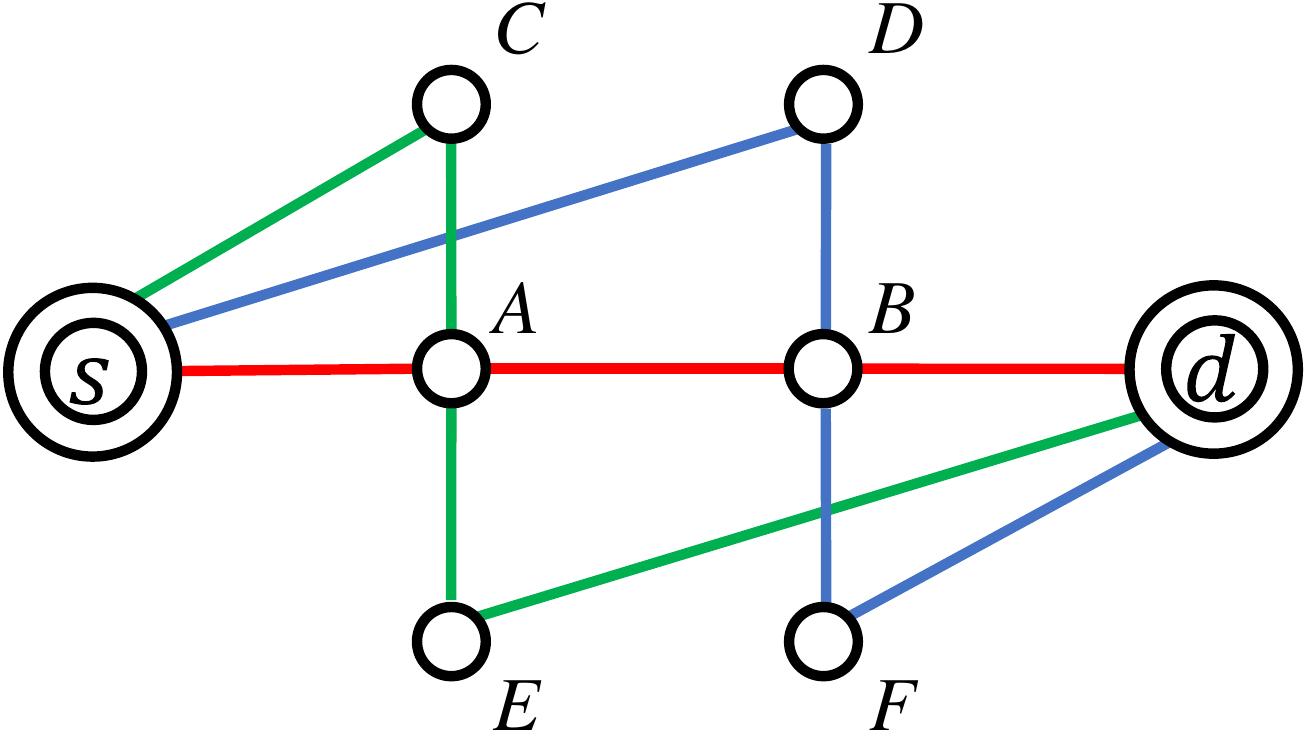}

		\caption{\small Counterexample for two possible algorithms}
		\label{fig:counterexample}

\end{figure}

We summarize the hardness of the multiple contention-free path selection problem without classifying it into a certain complexity class, and show its hardness in three examples. On one hand, because of the resource constraints (qubits/channels), path selection depends highly on the link states and hence the search space is much more than the classical algorithms which only depends on the weighted graph while edges and nodes have unlimited capacity; on the other hand, $E_t$ is non-linear, which invalids many existing proofs based on the linear additivity of the routing metric and thus degrades the efficiency of classical algorithms.

\textbf{Example 1.} Despite its good performance (shown in \S~\ref{sec:eva}), we prove G-EDA is not the optimal. An example graph is shown in Fig.~\ref{fig:counterexample} \footnote{Red path: $(s,A,B,d)$. Green path: $(s,C,A,E,d)$. Blue path: $(s,D,B,F,d)$.}. Suppose all the edges have width 3, all channels have creation rate  $p=0.99$, the swapping success rate $q=1$, $s$ and $d$ have qubit capacity 6, and all other nodes have capacity 3. Then the optimal contention-free paths are the blue path plus the green path. But the G-EDA will output only the red path. The reason of the failure of G-EDA is it falls in a local minimum and fails to give the max-flow -- the width of the red path is 3, as opposed to 6 for the blue path plus the green path.

\textbf{Example 2.} Though the classical max-flow algorithm gives the optimal solution, it performs worse than G-EDA in some other cases. Consider the same topology in Fig.~\ref{fig:counterexample} with changed parameters. Suppose all blue and green edges have width 1, red edges have width 2, all channels have creation rate  $p=0.6$, the swapping success rate $q=1$, $s$ and $d$ have qubit capacity 3, and all other nodes have capacity 2. From Fig.~\ref{fig:E-hops-0.6}, we know when $p=0.6$, one $(2,3)$-path is better than three $(1,4)$-paths. Hence, the optimal solution is the red path with $W=2$, which can be found via G-EDA. The max-flow algorithm, however, gives the green path, the blue path, and the red path -- all paths are single -- which is the sub-optimal solution.

\textbf{Example 3.} Due to the enormous search space, we failed to find the optimal strategy via brute-force even in a 10-node network. Suppose $|V|=10$, every node has 15 qubits and 6 edges, and each edge is composed of 5 quantum channels. In the brute-force searching, we do not assume the P2 and P4 are carried out based on `paths', but just try all possible assignments of qubits to channels, perform the swapping, calculate the $E_t$ between the given S-D pair, and record the highest result. For any S-D pair, the search space for P2 is $\sim 2.3 \cdot 10^{36}$\footnote{This number is got via a recursive algorithm instead of mathematical derivation. Consider the number of unique combinations of 15 indistinguishable balls put into 6 different buckets, each with capacity 5. }. Even worse, the quantum swapping in P4 depends on local states, which is prohibitively hard to enumerate all possible swapping combinations.

\subsection{Pseudocode for P2 algorithm of Q-PASS}

The pseudocode for P2 algorithm of Q-PASS is shown in Alg.~\ref{alg:external-1}. We define a function $Width(\cdot, \cdot)$, whose input are a path and current qubit capacity of all nodes, and output is the width of the path.

{\small
	\begin{algorithm}[h]
		\DontPrintSemicolon
		\SetKwInOut{Input}{Input}\SetKwInOut{Output}{Output}
		\small \Input{$G=\langle V, E, C \rangle$, $O$, $P$}
		\small \tcp{$O$: list of S-D pairs}
		\small \tcp{$P$: mapping from any S-D pair to its offline paths}
		\small \Output{$\langle L_C, L_P \rangle$}
		\small \tcp{$L_C$: list of channels to assign qubits}
		\small\tcp{ $L_P$: ordered list of selected paths}

		\small $L_C \gets \varnothing$\;
		\small $L_P \gets \varnothing$\;

		\small $T_Q \gets$ \textnormal{a table to map a node $x$ to its qubit capacity $Q_x$} \;
		\small \textnormal{construct} $T_Q$ \textnormal{from current topology}\;
		\small $W \gets \varnothing$\; \tcp{empty table to map a path $p$ to its width $w_p$}
		\small $q \gets \varnothing$\; \tcp{empty priority queue of paths, sorted by routing metric}

		\For{$o \in O$} {
			\For{$p \in P[o]$}{
				\small 	$T_W[p] \gets Width(p, T_Q)$\;
				\small 	$m \gets$ \textnormal{routing metric of $p$ with width W[p]} \;
				\small 	$q.\textnormal{enqueue}(p, m)$\;
			}
		}

		\While{$q$ \textnormal{is not empty}} {
			\small $p \gets q.\textnormal{dequeue}()$\;

			\If{$Width(p, T_Q) < width[p]$ } {
				\small \tcp{The width of $p$ has changed}
				\small \textnormal{Update $width[p]$ and re-insert $p$ to $q$} \;
				\small \textbf{continue}\;
			}

			\If{$Width(p, T_Q) = 0$ } {
				\small \tcp{Even the best path is unsatisfiable}
				\small \textbf{break}\;
			}

			$L_P \gets L_P + \langle p, width[p] \rangle$\;

			\For{$\langle n1, n2 \rangle \in $ \textnormal{edges of} $p$} {
				\small $T_Q[n1] \gets T_Q[n1] - width[p]$\;
				\small $T_Q[n2] \gets T_Q[n2] - width[p]$\;
				\small $L_C \gets L_C + $ \textnormal{$width[p]$ unbound channels on $\langle n1, n2 \rangle$} \;
			}
		}

		\small $partial \gets L_P + (q$ \textbf{as} List$)$ \;

		\For{$p \in partial$} {
			\small \textnormal{Update $T_Q$ and $L_C$ as line 21-23, only on available edges}\;
		}

		\caption{Adaptive resource allocation}
		\label{alg:external-1}
	\end{algorithm}
}

\subsection{Pseudocode for EDA}

We show the pseudocode of EDA in Alg.~\ref{alg:e-dijkstra}, and the time cost is analyzed in \S~\ref{sec:analysis}.

{\small
	\begin{algorithm}[h]
		\DontPrintSemicolon
		\SetKwInput{Input}{Input}
		\SetKwInput{Output}{Output}
		\small \Input{ $G=\langle V, E, C \rangle$, $e$, $\langle src, dst \rangle $ }
		\small \Output{ The best path $\langle p, W \rangle$}

		\SetKwInput{Proc}{Procedure}
		\small \tcp{Initialize empty states}
		\small $E \gets \textnormal{an array of } n \textnormal{elements, all set to} -\infty $ \;
		\small $prev \gets \textnormal{an array of } n \textnormal{ elements, all set to }\mathtt{null}$ \;
		\small $visited \gets \textnormal{an array of } n \textnormal{ elements, all set to } \mathtt{false}$ \;
		\small $width \gets \textnormal{an array of } n \textnormal{ elements, all set to } 0$ \;
		\small $q \gets \textnormal{fibonacci-heap, highest } E[\cdot] \textnormal{ first} $ \;

		\small \tcp{Initialize states of $src$}
		\small $E[src] \gets +\infty$ \;
		\small $width[src] \gets +\infty$ \;
		\small $q.\textnormal{enqueue}(src)$ \;

		\small \tcp{Track the best path until $dst$}
		\While { $q$ \textnormal{ is not empty }} {
			\small \tcp{Get the current best end node }
			\small $u \gets q.\textnormal{dequeue}()$ \;
			\small \lIf {$visited[u]$} {\textbf{continue}} \lElse {$visited[u] \gets \mathtt{true}$}

			\small \If {$u = dst$} {
				\small $\langle p, W \rangle \gets $ Construct path via $prev$ and $width$\;
				\small \textbf{return} $\langle p, W \rangle$ \;
			}

			\small \tcp{Expand one hop based on $u$}
			\small \For { $v \in$ \textnormal{neighbors of} $u$} {
				\small \lIf {$visited[v]$} {\textbf{continue}}

				\small $\langle p, W \rangle \gets $ Construct path via $prev$ and $width$\;

				\small $E' \gets e(p, W)$ \;

				\small \If {$E[v] < E'$} {
					\small 	$E[v] \gets E $ \;
					\small 	$prev[v] \gets u$ \;
					\small 	$width[v] \gets W$ \;
					\small 	$q.\textnormal{reorder}(v)$ \;
				}
			}
		}
		\caption{The Extended Dijkstra's algorithm \label{alg:e-dijkstra}}
	\end{algorithm}
}

\subsection{Time and space cost analysis}
\label{sec:analysis}

To avoid unbound computation and space cost in P2, we set the maximum number of multipath $K_m=200$. We set the maximum path hopcount according to the network itself. For any input $G$, 100 S-D pairs are randomly selected, and then multipath routing is performed via G-EDA between each S-D pair. The largest hopcount of selected paths whose $E_t>1$ is the maximum hopcount $h_m$ of all selected paths. We denote the number of nodes as $n$, the number of S-D pairs as $m$, and the maximum width of paths as $W_m$, which is determined by node capacities and edge widths.

\subsubsection{Cost of routing metric evaluation}

The calculation of $E_t$ can be performed by following the recursive formula set \ref{eq:P}. For an $h$-hop path with width $W$, the calculation of $E_t$ goes as following. Iterate on $k$, from 1 to $h$ and further iterate on $i$, from $W$ to $1$: calculate $Q_k^i$, $\sum_{l=i}^{W} Q_k^l$, $P_{k-1}^i$, $\sum_{l=i+1}^{W} P_{k-1}^l$, and $P_k^i$. Five $W$-element arrays are allocated to store the values. After that, $E_t = q^h \cdot \sum_{i=1}^{W} i \cdot P_h^i$ is calculated in $W+h$ time. Hence, the time cost is $O(hW)$, and the space cost is $O(W)$.

\subsubsection{Cost of P2 algorithm of Q-PASS}

The initialization costs $O(n)$ time. The double-\textbf{for} loop costs $O(m K_m(h_m+h_m+\log(m K_m)))$ time. The \textbf{while} loop costs $O(m K_m(h_m+\log(m K_m)+h_m))$ time. Hence, the overall time cost is $O(m K_m(h_m+\log(m K_m)))$.

Each of the $L_C$, $L_P$, and $q$ costs $O(m K_m h_m)$ space, the $T_Q$ costs $O(n)$ space, the $width$ costs $O(m K_m)$ space. Hence, the overall space cost is $O(m K_m h_m+n)$.

\subsubsection{Cost of EDA}

For a classical network $\langle V, E \rangle$, the Dijkstra's algorithm costs $O(n\log n + |E|)$ time because the dequeue operation costs $O(\log n)$ time, the reorder operation of the Fibonacci heap costs $O(1)$ time, and each edge is visited at most once. Similarly, the time cost for EDA is $O(n\log n + h_m W_m + |E|(h_m W_m))=O(n\log n + |E|(h_m W_m))$. The space cost is $O(n)$.

\end{document}